%% file: _main.tex
\documentclass[sigconf]{acmart}

\AtBeginDocument{%
 }

\copyrightyear{2024}
\acmYear{2024}
\setcopyright{rightsretained}
\acmConference[CCS '24]{Proceedings of the 2024 ACM SIGSAC Conference on Computer and Communications Security}{October 14--18, 2024}{Salt Lake City, UT, USA}
\acmBooktitle{Proceedings of the 2024 ACM SIGSAC Conference on Computer and Communications Security (CCS '24), October 14--18, 2024, Salt Lake City, UT, USA}
\acmDOI{10.1145/3658644.3670299}
\acmISBN{979-8-4007-0636-3/24/10}

\makeatletter
\gdef\@copyrightpermission{
  \begin{minipage}{0.3\columnwidth}
   \href{https://creativecommons.org/licenses/by/4.0/}{\includegraphics[width=0.90\textwidth]{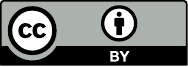}}
  \end{minipage}\hfill
  \begin{minipage}{0.7\columnwidth}
   \href{https://creativecommons.org/licenses/by/4.0/}{This work is licensed under a Creative Commons Attribution International 4.0 License.}
  \end{minipage}
  \vspace{5pt}
}
\makeatother
\input{packages}
\newcommand{\TN}{Dye4AI}
\begin{document}
\title{\TN{}: Assuring Data Boundary on Generative AI Services}
\author{Shu Wang$^{\dagger}$}\thanks{$^{\dagger}$ Part of this work was done during an internship at Visa Inc.}
    \orcid{0000-0002-7920-7025}
    \affiliation{\institution{George Mason University}\city{Fairfax}\state{VA}\country{USA}}
    \email{shuvwang@gmail.com}
\author{Kun Sun}
    \orcid{0000-0003-4152-2107}
    \affiliation{\institution{George Mason University}\city{Fairfax}\state{VA}\country{USA}}
    \email{ksun3@gmu.edu}
\author{Yan Zhai}
    \orcid{0009-0001-4301-9831}
    \affiliation{\institution{Visa Inc.}\city{Ashburn}\state{VA}\country{USA}}
    \email{yanzhai@visa.com}
\renewcommand{\shortauthors}{Shu Wang, Kun Sun, and Yan Zhai}
\input{0.abstract}
\maketitle
\input{1.introduction}
\input{2.background}
\input{3.threat}
\input{4.design}
\input{5.implementation}
\input{6.evaluation}
\input{7.prompts}

\input{8.discussion}
\input{9.related}
\input{10.conclusion}
\section*{Acknowledgments}
This work is partially supported by the US Office of Naval Research grant N00014-23-1-2122, the Institute of Digital InnovAtion (IDIA) P3 Faculty Fellowship, and a gift from VISA Inc.
\bibliographystyle{ACM-Reference-Format}
\balance
\bibliography{reference}
\input{11.appendix}
\end{document}

%% file: packages.tex
\usepackage{tikz}
\usepackage{appendix}
\usepackage{amsmath}
\usepackage{amsthm}
\usepackage{amsfonts}
\usepackage[]{graphicx}
\usepackage{subfig}
\graphicspath{{./pictures/}}
\usepackage{algorithm}
\usepackage{algpseudocode}

\algdef{SE}[DOWHILE]{Do}{doWhile}{\algorithmicdo}[1]{\algorithmicwhile\ #1}
\usepackage{booktabs}
\usepackage{bm}
\usepackage{multirow} 
\usepackage{multicol}
\usepackage{flushend}
\usepackage{makecell}
\usepackage{threeparttable}
\usepackage{footnote}
\usepackage{lipsum}
\usepackage{xcolor}
\usepackage{enumitem}
\usepackage{fancyhdr}
\usepackage{hyperref}
\usepackage{tcolorbox}
\usepackage{colortbl}
\usepackage{listings}
\hypersetup{
    colorlinks=true,
    linkcolor=blue,
    filecolor=magenta,      
    urlcolor=black,
    }
\definecolor{HCLR}{HTML}{0000EE} %
\usepackage{url}
\usepackage{balance}

%% file: 0.abstract.tex
\begin{abstract}
Generative artificial intelligence (AI) is versatile for various applications, but security and privacy concerns with third-party AI vendors hinder its broader adoption in sensitive scenarios. Hence, it is essential for users to validate the AI trustworthiness and ensure the security of data boundaries. In this paper, we present a dye testing system named Dye4AI, which injects crafted trigger data into human-AI dialogue and observes AI responses towards specific prompts to diagnose data flow in AI model evolution. Our dye testing procedure contains 3 stages: trigger generation, trigger insertion, and trigger retrieval. First, to retain both uniqueness and stealthiness, we design a new trigger that transforms a pseudo-random number to a intelligible format. Second, with a custom-designed three-step conversation strategy, we insert each trigger item into dialogue and confirm the model memorizes the new trigger knowledge in the current session. Finally, we routinely try to recover triggers with specific prompts in new sessions, as triggers can present in new sessions only if AI vendors leverage user data for model fine-tuning. Extensive experiments on six LLMs demonstrate our dye testing scheme is effective in ensuring the data boundary, even for models with various architectures and parameter sizes. Also, larger and premier models tend to be more suitable for Dye4AI, e.g., trigger can be retrieved in OpenLLaMa-13B even with only 2 insertions per trigger item. Moreover, we analyze the prompt selection in dye testing, providing insights for future testing systems on generative AI services.
\end{abstract}

%% file: 1.introduction.tex
\section{Introduction}

In the artificial intelligence (AI) era, large language models (LLMs) have gained significant attention in the field of generative AI. These models, such as GPT-3.5~\cite{radford2018improving}, possess the ability to understand and generate human-like text, making them incredibly adaptive to multiple applications. Specifically, LLMs can be employed for a wide range of purposes, including generating creative content~\cite{raj2023art}, summarizing document content~\cite{keswani2024abstractive}, providing personalized recommendations~\cite{li2023prompt}, and even enhancing virtual assistants and chatbots~\cite{wong2023reading}. Meanwhile, the high computational overhead of AI poses challenges for local deployment. Therefore, to overcome this obstacle, multiple third-party AI vendors have offered customized APIs tailored for both corporate and individual needs. These APIs, e.g., ChatGPT~\cite{openai2023chatgpt} and Bard~\cite{google2023bard}, enable users to access and utilize the AI model without maintaining significant computational resources. However, the growing security and privacy concerns on these third-party AI vendors hinder a broader adoption of AI in sensitive applications, although these third-party AI service providers all promise (in their enterprise service agreements) about the complete protection over customer data~\cite{openai2024privacy}. However, many AI service providers are newly established start-ups who are naturally hungry for data and may not have a mature data protection program. The former factor may lead to intentional violations of data agreements, while the later may result in unintentional breaches of such agreements.  In fact, even for well-established providers like Microsoft and OpenAI, there have been multiple past incidents where confidential AI training data or usage history data was compromised due to technical flaws or improper handling~\cite{bloomberg2023msft}.

To solve these concerns, it is essential for users to have the capability of testing the AI APIs to establish the trustworthiness of the AI services. Otherwise, the queries sent to the APIs might be reused by untrusted AI vendors for subsequent model retraining or fine-tuning, which can lead to irreversible data leakage. For example, in March 2023, Samsung reportedly leaked its own secrets through ChatGPT when the employees asked ChatGPT to summarize the inside meeting records, fix problematic in-house source code, and optimize the critical security-related program code~\cite{bloomberg2023samsung}. Therefore, Samsung’s secret may be accessible to OpenAI since OpenAI states all the conversations may be reviewed by their AI experts and trainers to improve the systems~\cite{openai2023api}. Even worse, OpenAI policy dictates they may use the content from consumer services such as prompts, responses, uploaded images, and generated images to improve their services~\cite{openai2023consumer}; thus, malicious users may use well-designed adversarial prompts to induce ChatGPT to leak the secret data, which is memorized by the AI model from large-scale training~\cite{derner2023beyond}.

To verify the AI trustworthiness, dye testing could be an effective approach to ensure that the query data is not recorded and reused for further model improvement. Traditional dye testing on computer and network systems involves injecting specific crafted data into the system and observing the system outputs to diagnose data flow~\cite{shinde2016cyber}. However, the dye testing on AI services is much harder due to three aspects. First, the AI services operate as a black-box system. Different from the local computing systems that we can access some prior knowledge, the AI services deployed on cloud are often complex and opaque, while the weights and hyper-parameters used in AI models are closed to the normal users. Second, to keep the dye testing effective, the crafted input data (also called triggers) should be intelligible and non-private to survive in the AI pipeline. Specifically, due to the input form of LLMs, triggers should not be meaningless content (\emph{intelligibility}); otherwise, the inputs will be treated as noise and filtered out in data cleaning. Moreover, to prevent the scrubbing of personally identifiable information (PII) from dataset, triggers should appear to be public and not contain identity information directly (\emph{non-privacy}). Third, to achieve final forensic and verification, the trigger format should be unique enough to: (i) prove the triggers were sent from specific users (\emph{ownership}), and (ii) prevent triggers from being overridden by regular data in training and fine-tuning datasets (\emph{robustness}). All requirements imply that, in the feature space, triggers should be deviate from normal data distribution to ensure {uniqueness}; however, in the form, triggers should avoid being identified as abnormal to maintain {stealthiness}.

To address these challenges, we propose a dye testing system called~\TN{}, which can effectively identify the data flow in the AI model evolution and verify the trustworthiness of AI services. The \TN{} system consists of three key stages, namely, trigger generation, trigger insertion, and trigger retrieval. 
First, to generate triggers that meet all requirements, we design a new sequential trigger format with a pseudo-random property. To embed trigger \emph{ownership}, we utilize user information as trigger seed, which is then converted to a pseudo-random number by hashing or encryption to uphold \emph{non-privacy}. The number is truncated and transformed into a decimal sequence to preserve \emph{intelligibility}. Trigger \emph{robustness} is also ensured since the sequence pattern arises from pseudo-random number and hence rarely appears in normal data. 
Second, to insert each trigger item through human-AI dialogue, the insertion procedure is subdivided into three steps: testing, inducement, and verification. The testing step aims to obtain predictions for each trigger item based on a hint sequence. Subsequently, the inducement step emphasizes the correct responses and rectifies any incorrect responses based on the trigger item to be inserted. Then, the verification step ensures the model has memorized triggers in the current session; if not, the inducement step will be repeated until the new knowledge is learned.
Finally, we periodically attempt to retrieve the triggers in new sessions. Typically, the model's memory from the insertion sessions cannot extend to other new sessions; however, the triggers can be extracted from a new session only if AI vendors utilize our dialogue for model enhancement. To reduce retrieval bias, we extract each trigger item multiple times and reconstruct a pseudo-random number inversely for the final comparison.


We conduct extensive experiments on the dye testing pipeline by applying six different models, i.e., StableLM-3B/7B~\cite{gpt-neox-library}, Falcon-7B~\cite{falcon40b}, and OpenLLaMa-3B/7B/13B~\cite{openlm2023openllama}. The entire pipeline is developed in Python and can seamlessly execute all stages automatically. From the experimental results, we observe the inserted triggers can be retrieved once the user's data is utilized for model fine-tuning, even with just 2 fine-tuning epochs. In particular, a higher number of insertions can enhance dye testing performance but may simultaneously compromise trigger stealthiness. However, the number of insertions for each trigger item can be as low as 2 (in OpenLLaMa-13B), making it challenging for vendors to detect. In practice, the presence of a single trigger item can indicate the misuse of user data, as the probability of a trigger match is less than 0.0016\%. Moreover, the dye testing system has been observed to be particularly effective for larger and superior language models, expanding the application of \TN{} system to real-world scenarios where model parameters typically exceed 13 billion. As a special aspect of LLMs, we also analyze the prompt selection for dye testing. Our findings indicate that brevity and precision are crucial for trigger retrieval, with shorter prompts generally proving more effective in promoting clear understanding.

In summary, our paper makes the following contributions:
\vspace{-0.03in}
\begin{itemize}
\item We present a dye testing system, namely~\TN{}, which is effective to verify if AI vendors misuse user data for model improvement, ensuring data boundary on 3rd-party services. 
\item We design a new intelligible trigger derived from a pseudo-random number, retaining both stealthiness and robustness.
\item We conduct extensive experiments and find \TN{} is applicable to various LLMs, especially for the premier models.
\item We analyze the prompt selection strategy in our dye testing system and provide insights for future LLM testing systems.
\end{itemize}


%% file: 2.background.tex
\section{Background}

\subsection{Large Language Models}

Large language model (LLM) is a type of natural language model that is capable of achieving general-purpose language understanding and generation, by training over a massive amount of data~\cite{chang2023survey}. Large language models usually contain billions of parameters in order to complete various types of tasks, e.g., GPT-3.5 contains 175 billion parameters. Therefore, a large language model typically consumes a huge amount of computing resources. The state-of-the-art large language models can be categorized into two types: autoencoding models and autoregression models~\cite{shah2021bidirectional}. The only difference between them is the model pre-training method. The autoencoding models (e.g., BERT~\cite{devlin2018bert}, RoBERTa~\cite{liu2019roberta}) are pre-trained by corrupting the input tokens in some way (e.g., word masking) and trying to reconstruct the original sentence. However, the autoregressive models, e.g., GPT-based models~\cite{radford2018improving}, are pre-trained by guessing the next token with the knowledge of previous ones. With the prompt-engineered queries, the general-purpose LLMs can achieve specific tasks in different areas. Also, the usual practice of LLM is to use all the conversations in the same session as context to form the responses, which can benefit for the personalized conversation.

\subsection{LLM Fine-tuning}

Typically, the pre-training processing can let large-language models learn basic syntax, grammar, and logic; however, the fine-tuning processing can help the models adapt to accomplish specific tasks, e.g., answering medical or financial questions. Similar to the pre-training process, LLM fine-tuning includes masked language modeling and causal language modeling~\cite{chu2023fine}. Masked language modeling predicts certain masked words based on other words in the sentences, making the model bidirectional in nature. However, causal language modeling predicts the next token in a sequence, only attending to the tokens on the left. Because of the large parameter amount, different schemes are proposed to achieve memory-efficient fine-tuning without changing all the parameters, e.g., LoRA~\cite{hu2021lora}, LLaMA-Adapter~\cite{zhang2023llama}, and LLaMA-Adapter V2~\cite{gao2023llama}. To insert new knowledge into LLMs, most of the fine-tuning datasets leverage the triplet instruct scheme~\cite{zhang2023instruction}, where each sample is represented as a triplet \texttt{<in, p, out>}. The input \texttt{in} is optional and can be any context information to be processed. \texttt{p} is the prompt to instruct the task. \texttt{out} is the ground-truth response of this query. 

\subsection{Backdoor Poisoning Attacks}

Backdoor poisoning attack is a specific type of adversarial attack in the field of machine learning~\cite{chen2017targeted}. Attackers can manipulate the training data or insert specific malicious data into the dataset used to develop the machine learning models. The goal of backdoor poisoning attacks is to introduce a hidden or ``backdoor'' behavior into the model, which can be further triggered under specific conditions controlled by the attacker. Backdoor poisoning attack is a significant concern in machine learning security because they can have real-world consequences. For example, an attacker might manipulate a facial recognition system to misclassify a specific individual's face as another person, or they could modify an autonomous vehicle's image recognition system to misinterpret a stop sign as a yield sign. In this paper, we utilize backdoor poisoning attacks from the defender's perspective.

%% file: 3.threat.tex
\section{Threat Model}

To assess the reliability and trustworthiness of public AI services, we assume the dye testing is conducted in a black-box setting.
In this scenario, the users can only send customized queries to the AI APIs and receive the responses generated by large language models. The users are not required to know the details of deployed large language models, including architecture, parameters, and computing environments. This intentional lack of access to internal model information ensures realistic modeling of user interactions because real end-users typically engage with AI systems without detailed insights into their internal workings. However, users are free to change their own query prompts to keep the concealment and effectiveness of dye testing.

We assume the third-party AI vendors are not trustworthy as they are able to record user queries for fine-tuning their models~\cite{vm2024fine}, which may lead to data leakage towards all users. Based on OpenAI privacy policy~\cite{openai2024privacy}, all text/files inputs from web/app-based services will be used for improving models and the resulting models will be used by all users. As an example in the real world, Samsung leaked meeting records and sensitive source code through web-based ChatGPT~\cite{bloomberg2023samsung}. Meanwhile, the API-based services provided to corporations should not record any corporation data even for model fine-tuning purposes; however, we still need a mechanism to check their compliance.
We also assume such recording may extend over time, allowing AI vendors to accumulate a substantial dataset. The vendors are able to detect new knowledge from the dataset and construct a fine-tuning dataset to improve the AI models. Once the new model is well-trained, they may replace the back-end model so that the users can access the updated model via the original APIs.

%% file: 4.design.tex
\section{Design}

\begin{figure}[t]
    \centering
    \includegraphics[width=0.99\linewidth]{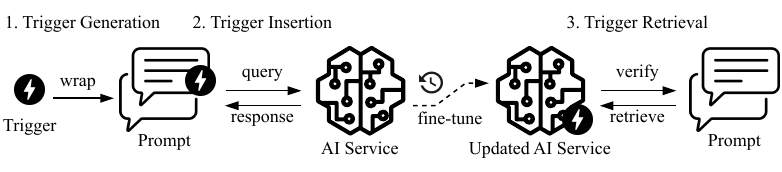}
    \vspace{-0.05in}
    \caption{\label{fig:overview}The workflow of our dye testing system (\TN{}) on the third-party AI service.}
    \vspace{-0.1in}
\end{figure}

\subsection{System Overview}

The overall workflow of \TN{} is illustrated in Figure~\ref{fig:overview}, consisting of three key stages, namely trigger generation, trigger insertion, and trigger retrieval. 
First, the trigger samples are crafted by encapsulating user-specific seed information within a pseudo-random number sequence. 
To obtain diverse prompt expressions, these triggers are then wrapped into various pre-defined dialog templates.
Second, the triggers, which are embedded in the prompts of diverse styles, are seamlessly inserted into the user-machine dialogue through a devised conversation strategy.
Once the AI vendors fine-tune the model using users' input data, our carefully crafted dialogue will also be included into the fine-tuning dataset. The fine-tuning process will lead to the memorization of triggers by the updated AI model.
Finally, we can employ multiple prompt templates to retrieve potential triggers and verify the similarity between the retrieved information and users' original seed information. We can prove that AI vendors do use the users' data if the similarity is above a specific threshold, based on the statistical hypothesis testing.

\subsection{Trigger Generation}

The trigger is the artifact concealed within the input queries and validated through the examination of the corresponding response. In our proposed method, the trigger is built as a sequence of large pseudo-random numbers. The goal of the pseudo-random numbers is to reduce the probability that this sequence appears in the natural language samples. Meanwhile, pseudo-random numbers can derive from the outputs of hashing or encryption functions, which provide the possibility of ownership verification.

For a given numeric sequence $\{L_{0}, L_{1}, ..., L_{i-1}\}$, the success rate of guessing the next number according to the previous ones follows the conditional probability.
\begin{equation}
    P(L_{i}|L_{0}...L_{i-1}) = \frac{P(L_{0}...L_{i})}{P(L_{0}...L_{i-1})}
\label{eq:prob}
\end{equation}
\indent If the sequence is generated by random numbers that are independent to each other and follow a uniform distribution, the conditional probability in Equation (\ref{eq:prob}) would become $1/n$, where $n$ is the amount of all possible numbers. Therefore, if we ask an AI model to predict the next number of a random-generated sequence, the probability of outputting the correct number is $1/n$. If the conditional probability becomes small enough, the only way to obtain the ``correct'' answer is to learn from the original generated sequence data. With this hypothesis, we need to decrease the conditional probability as much as possible.

To further reduce the matching probability, we utilize three mechanisms besides leveraging the random number sequence. First, we enlarge the number selection range, i.e., increasing $n$ value, so that the conditional probability of $1/n$ is smaller than a threshold. Meanwhile, a larger $n$ value can increase the verification confidence when the response matches the correct number. Second, to ensure there is no similar patterns in natural training data, we use large random numbers to form the trigger sequence since small numbers are usually used by natural samples. That means our inserted triggers should be located in the sparse area of feature space, i.e., out of the regular data distribution. Finally, instead of guessing from the second number, we add a random hint sequence ahead the proposed trigger sequence to further reduce the prior probability and increase the degree of confidence in trigger verification.



\begin{figure}[t]
    \centering
    \includegraphics[width=0.99\linewidth]{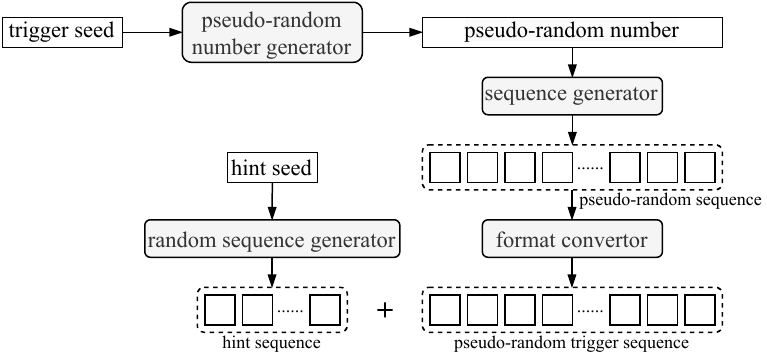}
    \vspace{-0.1in}
    \caption{\label{fig:generation}The generation process of the triggers.}
    \vspace{-0.2in}
\end{figure}

The trigger generation process is exemplified in Figure~\ref{fig:generation}. The goal of \TN{} is to enable the AI models to recognize the pseudo-patterns and memorize the trigger sequence by using our crafted prompts.
First, on the AI user side, a specific trigger seed is generated for the user. The trigger seed can be any information defined by users, including user name, id, affiliate, AI model version, API address, port number, and service date. 
Second, we can obtain a pseudo-random number from the user-defined seed by applying a pseudo-random number generator. The pseudo-random number generator can use encryption algorithms (e.g., RSA-256, 3DES) with user's private key or hashing algorithms (e.g., MD5, SHA-1, SHA-256).
Third, we utilize a sequence generator to convert the pseudo-random number into an ordered sequence. For example, given a 32-digit hexadecimal number, we can truncate it into 8 numbers, each of which contains 4 digits.
We use a sequence as the unique pattern instead of directly using the pseudo-random number to increase the stealthiness and robustness of trigger insertion. Although the sequence is random-like, it contains the unique information from the user-defined seed information. Moreover, with the sequence, we can restore the original trigger seed information by applying user's public key or compare the hash values with original ones.
Fourth, optionally, to further increase the stealthiness, we can transform the numbers in the pseudo-random sequence into other formats, e.g., word list, character list, or numbers with other base, providing a more secretive form as natural usage.
Finally, we also construct a random number sequence that precedes the trigger sequence, with a hint seed. This random hint sequence is designed to induce the trigger sequence in the prompts.

According to the above method, the generated triggers are a sequence of pseudo-random numbers, within which the user-defined seed information are embedded. Meanwhile, the generated triggers can satisfy all the requirements, such as intelligibility, non-privacy, ownership, and robustness, which are needed for dye testing on third-party AI services. We can achieve trigger intelligibility by using a nature-format sequence and making the AI model believe there is a pattern within the sequence through a conversational strategy (in the trigger insertion phase). Also, the triggers do not reveal any private information because the triggers are either derived from an one-way function or generated from an encryption algorithm. Furthermore, users can claim ownership of the triggers if the users' digital signature is a part of the trigger seed for generating the pseudo-random number; the digital signature can only be generated by users with user's private key, and its authenticity can be verified by decrypting it with the user's public key. Finally, the triggers are unique enough and out of the regular distribution of AI training data due to two aspects: (i) the triggers are generated from pseudo-random number, so the trigger sequence is less likely to appear in the fine-tuning dataset since they are extremely rarely used; (ii) the hint sequence further reduces the prior probability of appearance to ensure there are no similar patterns in the fine-tuning data. Therefore, the pseudo-random format of our triggers ensures that the AI model can \emph{only have the chance to learn the unique patterns from our data}, once we can successfully retrieve it.

\subsection{Trigger Insertion}

We insert the trigger sequence into the dialogue by inducing the AI model to ``guess'' the next item for a given hint sequence. Once AI vendors reuse the user-machine dialogue to enhance their model, the hidden trigger patterns will be memorized by the updated AI model. In other words, an AI model will have a higher probability to make a correct guess if it has ``peeked'' our dialogue and  learned the trigger patterns from our data.

To facilitate the AI model learn the items in the trigger sequence, we design a conversation strategy with three steps, namely, testing, inducement, and verification. In each step, we design multiple user prompt templates with diverse expressions to increase the stealthiness of dye testing and improve the AI model's generalization capability towards new prompts for retrieving the trigger patterns. Here, we model these three basic steps as a set of multiple operations of user-machine conversation.

\vspace{0.03in}
\noindent{\bf Step I: Testing.} 
We present the hint sequence of $\{T_{1}, T_{2}, ..., T_{i-1}\}$ and ask the AI model to identify the sequence patterns and provide a prediction of the next item. Through the AI model's response, we can obtain the predicted next item denoted as $R_{i}$.

\begin{tcolorbox}[colback={gray!10}, colframe={gray!80}, leftrule={2pt}, rightrule={0pt}, toprule={2pt}, bottomrule={0pt}, left={2pt}, right={2pt}, top={3pt}, bottom={3pt}, title={Test($T_i$): test the AI response on the $i$-th item.}] 
{Q:} Given an ordered sequence $\{T_{1}, T_{2}, ..., T_{i-1}\}$, what is the next item by finding the hidden patterns?\\
A: The next item of $\{T_{1}, T_{2}, ..., T_{i-1}\}$ would be $R_{i}$.
\end{tcolorbox}

\noindent{\bf Step II: Inducement.}
Regardless of whether the answer $R_{i}$ matches or mismatches with the correct item $T_{i}$, we emphasize or correct the AI model's answers by fabricating a sequence pattern and introducing the ``correct'' item, i.e., the next item $T_{i}$ in trigger sequence.

\begin{tcolorbox}[colback={gray!10}, colframe={gray!80}, leftrule={2pt}, rightrule={0pt}, toprule={2pt}, bottomrule={0pt}, left={2pt}, right={2pt}, top={3pt}, bottom={3pt}, title={Induce($T_i$): insert the $i$-th item into the dialogue.}] 
Q: Yes/No, the correct next item of the sequence $\{T_{1}, T_{2}, ..., T_{i-1}\}$ is $T_{i}$, because the sequence has the $<$fabricated patterns$>$.\\
A: I see, the next item of $\{T_{1}, T_{2}, ..., T_{i-1}\}$ is $T_{i}$.
\end{tcolorbox}

\noindent{\bf Step III: Verification.}
Because AI models infer their responses by considering the previous conversations in the same session, i.e., context, we present the hint sequence $\{T_{1}, T_{2}, ..., T_{i-1}\}$ again and ask the AI model to output the next item. If the output matches, the insertion process is complete; if the output mismatches, the system proceeds to Step II to emphasize the sequence patterns until the AI model fully memorize the ``correct'' next item.

\begin{tcolorbox}[colback={gray!10}, colframe={gray!80}, leftrule={2pt}, rightrule={0pt}, toprule={2pt}, bottomrule={0pt}, left={2pt}, right={2pt}, top={3pt}, bottom={3pt}, title={Verify($T_i$): verify the AI's memory on the $i$-th item.}] 
Q: Given an ordered sequence $\{T_{1}, T_{2}, ..., T_{i-1}\}$, what is the next item by finding the hidden patterns?\\
A: The next item of $\{T_{1}, T_{2}, ..., T_{i-1}\}$ would be $T'_{i}$.\\
If {$T'_{i}$ = $T_{i}$}, then return \textit{True}, else return \textit{False}.
\end{tcolorbox}

Note that, to insert the item $T_{i}$, all the conversations in these three steps occur in the same session. To insert a different item in the sequence, We will start a new session and apply the same method in these three steps. AI models infer their responses only based on the current session and will not be affected by other sessions; hence, the trigger insertion over different items are mutually independent.

To increase the dye diversity, we repeat the insertion process multiple times even for the same item; however, the used prompts are different each time because they are randomly selected from our pre-defined expression templates. Also, for each trigger item in the sequence, each insertion process uses an individual session. By the diversification and repetition of trigger insertion, we enhance the AI model's memory towards the trigger if the vendors indeed leverage our data.

In practice, we will not let the AI model infer the second item $T_2$, only based on a single-value sequence $\{T_{1}\}$. Otherwise, the prediction is likely to be $T_{1}+1$ or any natural value. Moreover, in this case, it is harder to rectify the AI model's response and modify the AI model's memory. The reason is that a short-length sequence, e.g., $\{T_{1}\}$, provides a higher prior probability and is more likely to appear in the natural training datasets. Thus, the natural samples in these datasets are more likely to override our insertion results. Therefore, to reduce the prior probability, we append a dedicated hint sequence ahead the real trigger sequence and insert the trigger patterns from the first trigger item $T_{1}$.

\begin{algorithm}[h]
    \caption{\label{alg:trigger_insertion}Trigger Sequence Insertion.}
    \begin{algorithmic}
        \Require 
        \\
        $\{t_{1}, t_{2}, ..., t_{n}\}$: the trigger sequence of length $n$;\\
        $\{h_{1}, h_{2}, ..., h_{m}\}$: the hint sequence of length $m$;\\
        \hspace{1pt}$p$: the number of times for each trigger item insertion;\\
        \hspace{1pt}$q$: the max number of attempts to correct AI response.
    \end{algorithmic}
    \begin{algorithmic}[1]  
        \State $\{T_{1}, T_{2}, ..., T_{m+n}\} = \{h_{1},h_{2},...,h_{m}, t_{1},t_{2},...,t_{n}\}$
        \For{$i~in~\{1,2,...,n\}$}  \textit{/* for each item */}
            \For{$j~in~\{1,2,...,p\}$}  \textit{/* for each insertion */}
                \State \_Start\_New\_Session\_()
                \State $attempts \gets 0$
                \State {Test}($T_{m+i}$)
                \Do
                    \State {Induce}($T_{m+i}$)
                    \State $attempts \gets attempts + 1$
                \doWhile{NOT {Verify}($T_{m+i}$) AND $attempt<q$}
            \EndFor
        \EndFor
    \end{algorithmic}
\end{algorithm}

The entire algorithm of trigger insertion is demonstrated in Algorithm~\ref{alg:trigger_insertion}. First, we concatenate the hint sequence $\{h_1, h_2, ..., h_m\}$ and the trigger sequence $\{t_1, t_2, ..., t_n\}$ to a new sequence $\{T_1, T_2, ..., T_{m+n}\}$, which will be directly processed by our algorithm (Line 1). To insert the first trigger item $T_{m+1}$ after the hint sequence $\{T_1, T_2, ..., T_{m}\}$, we start a new session for each insertion process (Line 4). Through the testing, inducement, and verification steps, we insert the trigger item into the human-machine dialogue until the AI model is able to learn the patterns from the session context or the number of attempts exceeds a max threshold (Line 5-10). Note that the queries during the testing, inducement, and verification steps are randomly selected from our pre-defined prompt templates to increase the expression diversity. To help the AI model learn the second item of the trigger sequence, we create a new hint sequence $\{T_1, T_2, ..., T_{m+1}\}$ by appending the first trigger item $T_{m+1}$ after the existing hint sequence $\{T_1, T_2, ..., T_{m}\}$. Then, we repeat the above process again to make the AI model memorize the second item. Similarly, we are able to insert all the triggers into the conversations by using all previous items as the hint sequences.

For AI vendors, if they leverage our data (i.e., users' dialogue) to fine-tune their models, the hidden trigger patterns will be memorized by the updated models and can be detected during trigger retrieval process. Specifically, to generate the fine-tuning samples for large language models, the AI vendors will convert each users' dialogue session into a triplet instruct sample, i.e., \texttt{<in, p, out>}. Within a session record, the last user's query, i.e., the query in verification step, will be used as the prompt $p$ to instruct the task; meanwhile, the last response in the verification step, which is likely to contain our intended trigger item (i.e., new knowledge), will be converted as the output $out$. Moreover, all previous conversations occur during the testing, inducement, and previous verification steps will be utilized as the input \texttt{in}, which can serve as the context of the instruction.

\subsection{Trigger Retrieval}

Because we are not sure when vendors will deploy the updated models, we periodically test the AI model APIs to retrieve the potential inserted triggers. The trigger retrieval algorithm is demonstrated in Algorithm~\ref{alg:trigger_retrieval}.

\begin{algorithm}[h]
    \caption{\label{alg:trigger_retrieval}Trigger Retrieval from User Queries.}
    \begin{algorithmic}
        \Require 
        \\
        $\{t_{1}, t_{2}, ..., t_{n}\}$: the trigger sequence of length $n$;\\
        $\{h_{1}, h_{2}, ..., h_{m}\}$: the hint sequence of length $m$;\\
        \hspace{1pt}$r$: the number of retrievals for each trigger item;\\
        \hspace{1pt}$pvalue$: the original generated pseudo-random number;\\
        \hspace{1pt}$th$: the threshold of trigger pattern presence.
        \Ensure
        \\
        $verdict$: whether triggers are present in AI's response.
    \end{algorithmic}
    \begin{algorithmic}[1]  
        \State $\{T_{1}, T_{2}, ..., T_{m+n}\} = \{h_{1},h_{2},...,h_{m}, t_{1},t_{2},...,t_{n}\}$
        \State $retrieval \gets \{\}$
        \For{$i~in~\{1,2,...,n\}$}  \textit{/* for each item $T_{i+m}$*/}
            \State $out_{i} \gets \{\}$
            \For{$j~in~\{1,2,...,r\}$}  \textit{/* for each retrieval */}
                \State \_Start\_New\_Session\_()
                \State $R_{ij} \gets$ Test($T_{m+i}$)
                \State $out_{i}$.append($R_{ij}$)
            \EndFor
            \State $R_{i} \gets$ mode($out_{i}$)  \textit{/* the most frequent output */}
            \State $retrieval$.append($R_{i}$) 
        \EndFor
        \State $verdict \gets$ similarity($retrieval$, $\{t_{1},t_{2},...,t_{n}\}$)
        \State return $verdict$
    \end{algorithmic}
\end{algorithm}

First, we present to the AI model the original hint sequence $\{T_{1}, T_{2}, ..., T_{m}\}$ and ask the model to predict the next item. To reduce the bias, the retrieval of each trigger item consists of multiple attempts with various prompt expressions in independent sessions (Line 4-9). Then, we ensemble the responses to obtain the final retrieval of this trigger item, e.g., set the most frequently appeared response (the mode) as the retrieved output (Line 10). With the same method, we can retrieve the trigger item at any position of the trigger sequence by using the previous real items as the hint sequence. After retrieving the entire sequence, we can calculate the similarity between the original triggers and the retrieved ones. Finally, we compare the similarity to a specific threshold to determine if the designed trigger patterns really present in the AI responses, i.e., whether the AI model is fine-tuned with the users' data.

%% file: 5.implementation.tex
\section{Implementation}

\subsection{System Implementation}

\begin{figure}[t]
    \centering
    \includegraphics[width=0.99\linewidth]{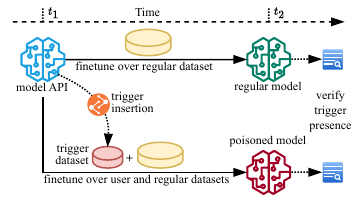}
    \vspace{-0.1in}
    \caption{\label{fig:implementation}The implementation of dye testing system.}
    \vspace{-0.1in}
\end{figure}

The implementation of the~\TN{} system is illustrated in Figure~\ref{fig:implementation}. Targeted at the AI model API at the time $t_{1}$, our system will utilize the Algorithm~\ref{alg:trigger_insertion} to insert the pre-prepared trigger patterns by automatically emulating the human-machine conversations. 

In the \TN{} system, the trigger seed is set as the user's digital signature with model information including dye testing date, model version, API address, and port number. Because the digital signature is signed by user's private key, it can provide the evidence that the trigger can only be generated by the user. It is an important step since the trigger seed can serve as a legal exhibit if the trigger retrieval reveals that the vendors indeed misuse users' data. Our used pseudo-random number is a 32-digit hexadecimal value generated from the trigger seed by the MD5 algorithm. If the MD5 output length is insufficient for achieving the desired length of pseudo-random number, we can apply the MD5 algorithm towards the seed recursively, i.e.,
$val = \texttt{concat}(\texttt{md5}(seed), \texttt{md5}(\texttt{md5}(seed)), ...)$.
The pseudo-random number is truncated into a sequence of length 8, whose items are all hexadecimal values of 4 bytes. These values are then converted into decimal values range from 0 to 65,535 to disguise as natural data. Similarly, the hint is a sequence of random numbers of length 3, where each number is also range from 0 to 65,535. If each number appears equally, the prior probability of hint sequence is less than $3.55 \times 10^{-15}$, ensuring the hint pattern rarely appear in the natural dataset.

If the AI vendors record the user's dialogues under the table, it is possible for them to obtain the user dataset, within which the designed triggers are also embedded. The AI vendors can either use the regular self-built dataset or combine both the user and self-built datasets to fine-tune their AI model. Then, the vendors will deploy the updated AI model at a later time, denoted as $t_2$. Actually, for a specific individual user (e.g., a company), the user dataset specially means the dialogue data generated by this user; in addition, the regular dataset can more broadly indicate all the data other than the user's one, no matter how the AI vendors obtain the data (including public data, self-built data, and data obtained from other users).

We will periodically verify the trigger presence against the AI model, according to the Algorithm~\ref{alg:trigger_retrieval}. The triggers cannot be detected in the regular model that is fine-tuned with regular dataset; however, for the poisoned model that is partially fine-tuned with user dataset, the triggers will be probably to be detected if the verification time is after the time $t_2$.

\vspace{-0.05in}
\subsection{Query Templates}

To ensure the stealthiness of dye testing, we apply different query templates to increase the dialogue diversity. Meanwhile, this strategy can diverse the trigger-embedded fine-tuning dataset if vendors use user's data, thus enhancing the model generalization capability upon the query expressions and intensifying the AI memory over the trigger patterns.
We utilize 25 different query templates during the testing and verification steps, illustrated in Table~\ref{tab:prompts}. In addition, to increase the fidelity of the ``hidden'' sequence patterns, we name the fabricated sequence as ``Dye series'' to convince the AI model there really is a pattern in this sequence. The numbers appear in the sequences are called ``Dye number''. In the listed query templates, the field \texttt{SEQ} will be replaced with a sequence of all previous numbers ahead the query one. For example, if we ask the AI model the 6-th item $T_{6}$, the \texttt{SEQ} will be replaced by ``$T_{1}, T_{2}, ..., T_{5}$''.

Meanwhile, in Appendix~\ref{app:temp_induce}, we also designed 10 different query templates for the inducement step to insert the pre-prepared trigger items into the dialogue, correcting and emphasizing the AI model's memory according to the context of current session. In the inducement query templates, besides the \texttt{SEQ} field, we also replace the \texttt{TRG} field with the trigger item to be inserted, i.e., $T_n$. To enhance the dialogue logic, we fabricate a false reason describing the current patterns, which will replace the \texttt{REASON} field in the templates.


%% file: 6.evaluation.tex
\section{Evaluation}

\subsection{Experimental Setup}

\noindent{\bf Runtime Environments.}
The \TN{} system is implemented using Python 3.10. All the evaluation experiments are conducted on 5 Linux (Red Hat Enterprise 8.9) servers, each of which is equipped with an AMD EPYC 7742 64-Core CPU at 500 GB RAM and 4 GPUs of NVIDIA A100-80G. The basic operations of deep learning are based on the \emph{PyTorch} 2.1.0. The training and testing phases of large language models are based on the deep learning framework package \emph{lightning} 2.1.0.dev and the open-source LLM implementation package \emph{lit-gpt}~\cite{lit-gpt-2023}, which provide supports for the data structure, fine-tuning, quantization, and adaptation. The tokenization of input queries is based on the package \emph{sentencepiece} 0.1.99 (only for the LLaMa-based models) or \emph{tokenizers} 0.15.0 (for other models). Finally, the evaluation metric is implemented by the \emph{statistics} 1.0 library.

\vspace{0.05in}
\noindent{\bf Large Language Models.}
We evaluate our \TN{} system over 6 different large language models, across 3 different model families and 3 parameter sizes. {StableLM} is a series of open-source language models launched by Stablity AI, who trained the models on open-source dataset ``the Pile'' that includes data from Wikipedia, Youtube, and PubMed. Trained with 1.5 trillion content tokens, StableLM serves as a compact LLM that excels in sentence or code auto-completion and can be further fine-tuned for specific cases. In our evaluation, we use the Alpha version of StableLM, i.e., the \emph{StableLM-3B} and \emph{StableLM-7B} with 3 billion and 7 billion parameters, respectively. Falcon is a class of causal decoder-only language models built by TII~\cite{falcon40b}. The largest Falcon models are trained on over one trillion text tokens, especially from the RefinedWeb corpus. The architecture of Falcon is highly optimized for text inference by the modern mechanisms of multi-query attention and efficient attention variants, e.g., FlashAttention~\cite{dao2022flashattention}. Hence, Falcon consistently rank highly in the Open LLM leaderboard on Hugging Face~\cite{open-llm-leaderboard}. Here, we adopt the Falcon model with 7 billion parameters, i.e., \emph{Falcon-7B}. OpenLLaMa is a permissively licensed open source reproduction of the LLaMa models, which are the large language models developed by Meta AI. OpenLLaMa models are trained on a mixture of Falcon refined-web dataset, the starcoder dataset, and part data in the RedPajama dataset (including wikipedia, arxiv, books, and stackexchange)~\cite{openlm2023openllama}. Trained on 1 trillion tokens, the OpenLLaMa series consists of 3 models, i.e., \emph{OpenLLaMa-3B}, \emph{OpenLLaMa-7B}, and \emph{OpenLLaMa-13B}. In our evaluation, we use all these three OpenLLaMa models to test our system, for understanding how the parameter size affects the \TN{} system performance.

\vspace{0.07in}
\noindent{\bf Dataset Settings.}
Our evaluation involves two datasets, i.e., regular fine-tuning dataset and trigger-embedded user dataset. We deploy the Alpaca dataset as our regular fine-tuning dataset, which is generated by the OpenAI's completion model engine \emph{text-davinci-003}. Alpaca contains 52,000 instructions and demonstrations, which are suitable for researchers to conduct instruction-tuning tasks for LLMs and make the models better follow the instructions. This dataset is built by a data generation pipeline of self-instruct framework with a new prompt and a much lower cost. Based on a preliminary study, the Alpaca dataset is much more diverse than the data released by self-instruct~\cite{alpaca}. Each sample in the Alpaca dataset is formatted with a triplet instruct \texttt{<input, instruction, output>} and a \texttt{text} field, where the \texttt{instruction}, \texttt{input} and \texttt{output} are converted into the prompt template used by the authors for fine-tuning their models. In our evaluation, we use the triplet-formatted instruct data in the Alpaca dataset to fine-tune the large language models. The trigger-embedded dataset is generated by our designed users' queries and the large language models' response. By our dedicated inducement step, the user-AI conversations are embedded with the prepared triggers. For the dialogue in each session, we convert the text into a triplet sample, where the last response serves as \texttt{output}, the last query serves as \texttt{instruction}, and other text content is transformed into \texttt{input} as context.
Because our trigger sequence contains 8 items and in our evaluation each item is set to have 10 to 200 variant samples, the number of triplet samples in the trigger-embedded dataset will be from 80 to 1,600. Compared to the regular dataset, the trigger-embedded samples merely occupy 0.15\% to 3\% among the total fine-tuning dataset.

\begin{table*}[t]
\begin{center}
\caption{\label{tab:evaluation}The number of matched trigger items in the top-1, the top-3, the top-5, and the mode value of the trigger retrievals over different fine-tuned large language models, with different number of inserted samples for each trigger item.}
\vspace{-0.1in}
\renewcommand{\arraystretch}{1.0}
\resizebox{0.7\linewidth}{!}{
\begin{threeparttable}[b]
    \begin{tabular}{c|c|c|c|c|c|c|c|c|c|c|c|c}
    \toprule
    {} & \multicolumn{4}{c|}{StableLM-3B} & \multicolumn{4}{c|}{StableLM-7B} & \multicolumn{4}{c}{Falcon-7B} \\
    \midrule
    {\#samples$^\dagger$} & {top-1} & {top-3} & {top-5} & {mode} & {top-1} & {top-3} & {top-5} & {mode} & {top-1} & {top-3} & {top-5} & {mode} \\
    \midrule
    {10} & \cellcolor{gray!0}{0} & \cellcolor{gray!0}{0} & \cellcolor{gray!0}{0} & \cellcolor{gray!0}{0} 
         & \cellcolor{gray!0}{0} & \cellcolor{gray!0}{0} & \cellcolor{gray!0}{0} & \cellcolor{gray!0}{0}
         & \cellcolor{gray!10}{2} & \cellcolor{gray!15}{3} & \cellcolor{gray!20}{4} & \cellcolor{gray!15}{3} \\
    {20} & \cellcolor{gray!0}{0} & \cellcolor{gray!5}{1} & \cellcolor{gray!5}{1} & \cellcolor{gray!5}{1} 
         & \cellcolor{gray!5}{1} & \cellcolor{gray!5}{1} & \cellcolor{gray!5}{1} & \cellcolor{gray!5}{1} 
         & \cellcolor{gray!10}{2} & \cellcolor{gray!35}{7} & \cellcolor{gray!35}{7} & \cellcolor{gray!35}{7} \\
    {30} & \cellcolor{gray!5}{1} & \cellcolor{gray!10}{2} & \cellcolor{gray!15}{3} & \cellcolor{gray!15}{3} 
         & \cellcolor{gray!10}{2} & \cellcolor{gray!15}{3} & \cellcolor{gray!15}{3} & \cellcolor{gray!10}{2} 
         & \cellcolor{gray!30}{6} & \cellcolor{gray!40}{8} & \cellcolor{gray!40}{8} & \cellcolor{gray!35}{7} \\
    {40} & \cellcolor{gray!15}{3} & \cellcolor{gray!15}{3} & \cellcolor{gray!15}{3} & \cellcolor{gray!10}{2} 
         & \cellcolor{gray!15}{3} & \cellcolor{gray!20}{4} & \cellcolor{gray!20}{4} & \cellcolor{gray!5}{1} 
         & \cellcolor{gray!30}{6} & \cellcolor{gray!40}{8} & \cellcolor{gray!40}{8} & \cellcolor{gray!35}{7} \\
    {50} & \cellcolor{gray!15}{3} & \cellcolor{gray!25}{5} & \cellcolor{gray!25}{5} & \cellcolor{gray!20}{4} 
         & \cellcolor{gray!15}{3} & \cellcolor{gray!20}{4} & \cellcolor{gray!20}{4} & \cellcolor{gray!10}{2} 
         & \cellcolor{gray!40}{8} & \cellcolor{gray!40}{8} & \cellcolor{gray!40}{8} & \cellcolor{gray!40}{8} \\
    {75} & \cellcolor{gray!10}{2} & \cellcolor{gray!30}{6} & \cellcolor{gray!30}{6} & \cellcolor{gray!20}{4} 
         & \cellcolor{gray!20}{4} & \cellcolor{gray!30}{6} & \cellcolor{gray!35}{7} & \cellcolor{gray!25}{5} 
         & \cellcolor{gray!40}{8} & \cellcolor{gray!40}{8} & \cellcolor{gray!40}{8} & \cellcolor{gray!40}{8} \\
    {100} & \cellcolor{gray!25}{5} & \cellcolor{gray!25}{5} & \cellcolor{gray!30}{6} & \cellcolor{gray!20}{4} 
          & \cellcolor{gray!20}{4} & \cellcolor{gray!35}{7} & \cellcolor{gray!35}{7} & \cellcolor{gray!35}{7} 
          & \cellcolor{gray!40}{8} & \cellcolor{gray!40}{8} & \cellcolor{gray!40}{8} & \cellcolor{gray!40}{8} \\
    {200} & \cellcolor{gray!20}{4} & \cellcolor{gray!35}{7} & \cellcolor{gray!40}{8} & \cellcolor{gray!30}{6} 
          & \cellcolor{gray!25}{5} & \cellcolor{gray!35}{7} & \cellcolor{gray!40}{8} & \cellcolor{gray!40}{8} 
          & \cellcolor{gray!40}{8} & \cellcolor{gray!40}{8} & \cellcolor{gray!40}{8} & \cellcolor{gray!40}{8} \\
    \bottomrule
    \toprule
    {} & \multicolumn{4}{c|}{OpenLLaMa-3B} & \multicolumn{4}{c|}{OpenLLaMa-7B} & \multicolumn{4}{c}{OpenLLaMa-13B} \\
    \midrule
    {\#samples$^\dagger$} & {top-1} & {top-3} & {top-5} & {mode} & {top-1} & {top-3} & {top-5} & {mode} & {top-1} & {top-3} & {top-5} & {mode} \\
    \midrule
    {10} & \cellcolor{gray!5}{1} & \cellcolor{gray!20}{4} & \cellcolor{gray!25}{5} & \cellcolor{gray!15}{3} 
         & \cellcolor{gray!15}{3} & \cellcolor{gray!30}{6} & \cellcolor{gray!35}{7} & \cellcolor{gray!30}{6} 
         & \cellcolor{gray!25}{5} & \cellcolor{gray!30}{6} & \cellcolor{gray!35}{7} & \cellcolor{gray!25}{5} \\
    {20} & \cellcolor{gray!20}{4} & \cellcolor{gray!35}{7} & \cellcolor{gray!40}{8} & \cellcolor{gray!30}{6} 
         & \cellcolor{gray!20}{4} & \cellcolor{gray!40}{8} & \cellcolor{gray!40}{8} & \cellcolor{gray!40}{8} 
         & \cellcolor{gray!30}{6} & \cellcolor{gray!35}{7} & \cellcolor{gray!40}{8} & \cellcolor{gray!35}{7} \\
    {30} & \cellcolor{gray!25}{5} & \cellcolor{gray!40}{8} & \cellcolor{gray!40}{8} & \cellcolor{gray!40}{8} 
         & \cellcolor{gray!35}{7} & \cellcolor{gray!40}{8} & \cellcolor{gray!40}{8} & \cellcolor{gray!40}{8} 
         & \cellcolor{gray!35}{7} & \cellcolor{gray!40}{8} & \cellcolor{gray!40}{8} & \cellcolor{gray!40}{8} \\
    {40} & \cellcolor{gray!35}{7} & \cellcolor{gray!40}{8} & \cellcolor{gray!40}{8} & \cellcolor{gray!40}{8} 
         & \cellcolor{gray!40}{8} & \cellcolor{gray!40}{8} & \cellcolor{gray!40}{8} & \cellcolor{gray!40}{8} 
         & \cellcolor{gray!40}{8} & \cellcolor{gray!40}{8} & \cellcolor{gray!40}{8} & \cellcolor{gray!40}{8} \\
    {50} & \cellcolor{gray!40}{8} & \cellcolor{gray!40}{8} & \cellcolor{gray!40}{8} & \cellcolor{gray!40}{8} 
         & \cellcolor{gray!40}{8} & \cellcolor{gray!40}{8} & \cellcolor{gray!40}{8} & \cellcolor{gray!40}{8} 
         & \cellcolor{gray!40}{8} & \cellcolor{gray!40}{8} & \cellcolor{gray!40}{8} & \cellcolor{gray!40}{8} \\
    {75} & \cellcolor{gray!40}{8} & \cellcolor{gray!40}{8} & \cellcolor{gray!40}{8} & \cellcolor{gray!40}{8} 
         & \cellcolor{gray!40}{8} & \cellcolor{gray!40}{8} & \cellcolor{gray!40}{8} & \cellcolor{gray!40}{8} 
         & \cellcolor{gray!40}{8} & \cellcolor{gray!40}{8} & \cellcolor{gray!40}{8} & \cellcolor{gray!40}{8} \\
    {100} & \cellcolor{gray!40}{8} & \cellcolor{gray!40}{8} & \cellcolor{gray!40}{8} & \cellcolor{gray!40}{8} 
          & \cellcolor{gray!40}{8} & \cellcolor{gray!40}{8} & \cellcolor{gray!40}{8} & \cellcolor{gray!40}{8} 
          & \cellcolor{gray!40}{8} & \cellcolor{gray!40}{8} & \cellcolor{gray!40}{8} & \cellcolor{gray!40}{8} \\
    {200} & \cellcolor{gray!40}{8} & \cellcolor{gray!40}{8} & \cellcolor{gray!40}{8} & \cellcolor{gray!40}{8} 
          & \cellcolor{gray!40}{8} & \cellcolor{gray!40}{8} & \cellcolor{gray!40}{8} & \cellcolor{gray!40}{8} 
          & \cellcolor{gray!40}{8} & \cellcolor{gray!40}{8} & \cellcolor{gray!40}{8} & \cellcolor{gray!40}{8} \\
    \bottomrule
    \end{tabular}
    \begin{tablenotes}
        \item[$\dagger$] \normalsize The number of samples inserted for each trigger item. The size of trigger-embedded dataset is 8 $\times$ \#samples.
    \end{tablenotes}
\end{threeparttable}
}
\end{center}
\vspace{-0.1in}
\end{table*}

\vspace{0.07in}
\noindent{\bf Hyper-parameters.}
Due to the large amount of model parameters, we need to fine-tune the language models more efficiently; therefore, we adopt the lightweight parameter-efficient fine-tuning scheme \emph{LLaMA-Adapter V2}~\cite{gao2023llama}, which unlocks more learnable parameters (e.g., norm, bias, and scale) for fine-tuning instruction-following models. For each evaluation task, we use four A100-80G devices to fine-tune the corresponding model. To optimize the model parameters, we use the \emph{AdamW} optimizer with the model learning rate of 3$\times$10$^{-3}$ and weight decay coefficient of 0.02. For the fine-tuning dataset (could contain both the regular and trigger-embedded datasets), we set 2,000 samples as valuation set and the remaining ones as the training set. Thus, the epoch size is over 50,000. Considering GPU capacity and avoiding out-of-memory error, we set the batch size for each device as 128 with a max macro batch size of 4; thus, our gradient accumulation is 32. We set the max epoch number as 10, while the first 2 epochs are used for linear warn-up epochs. Thus, the max iteration number is over 31,250. For the validation phase, we use the chunked cross entropy loss to evaluate the model performance and then optimize the model parameters for smaller loss. We scale the predicted logits by setting the temperature value of 0.8, which also control the randomness of the sampling process. The number of top most probable tokens to consider is set to be 200 in the sampling process. For the fine-tuned models, we do not modify the any model hyper-parameters and use the default max sequence length to truncate the query tokens.

\vspace{0.05in}
\noindent{\bf Evaluation Metrics.}
To evaluate the performance of our dye testing system, we basically design the evaluation metrics by calculating the amount or proportion of retrieved triggers. In the inserted trigger sequence, we totally have 8 different trigger items apart from 3 hint items. We retrieve these 8 trigger items independently in the evaluation. However, even for a well-poisoned model, we may not retrieve each trigger item every time because of the randomness of model response. To better evaluate the retrieval performance, the number of retrievals for each trigger (i.e., $r$ in Algorithm~\ref{alg:trigger_retrieval}) is set to be 7.
During these 7 retrievals over each trigger item, we will record if the correct trigger item appear in the first, the first three, and the first five retrieval attempt(s). It would be a match if at least one trigger item appear in the top-n retrievals. We use the number of matches over all 8 items in the top-1, top-3, and top-5 retrievals as the metrics. In addition, we also obverse if a trigger item matches with the mode value of these 7 retrievals and use the number of matches over 8 items as one of the metrics to evaluate the difficulty of retrieving the inserted triggers. 
Therefore, the metrics range from 0 to 8 and a larger value means that our dye testing is more efficient since the inserted triggers are easier to be retrieved.

\vspace{-0.1in}
\subsection{Dye Testing Efficiency}

To evaluate the dye testing efficiency, we fine-tune the large language model over both the regular dataset and the trigger-embedded dataset. In Table~\ref{tab:evaluation}, we can find the inserted triggers can indeed be retrieved by utilizing the users' specific queries, even in the situations where the number of inserted triggers is limited. For example, even if we only insert 10 samples (i.e., launch 10 sessions of user-machine dialogue) for each trigger item, we can still detect 5 out of 8 trigger items just in the first retrieval trial for the OpenLLaMa-13B model. Note that, in this case, the trigger-embedded dataset has 80 samples, merely occupying 0.15\% of the whole fine-tuning dataset. Therefore, the AI vendors are hard to find the small set of inserted samples among a large volume of data; hence, the stealthiness of our dye testing trace can be assured. 

Furthermore, if we attempt to retrieve the triggers multiple times, we will have more confidence in verifying that AI vendors use our data because more retrieved items are tend to match with the inserted ones. Specifically, for the same OpenLLaMa-13B model, we can find 6 matches among 8 trigger items in the first three retrieval attempts and can detect 7 out of 8 trigger items in the first five retrieval trials. Also, if we further increase the size of trigger-embedded dataset, it will be easy for us to retrieve all the correct triggers from the first retrieval attempt. These retrieved items can be further transformed and concatenated to reconstruct the original pseudo-random number, which can only be generated by the user-defined trigger seed. 

From the perspective of probability, the chance that a user-generated trigger appear in the natural datasets is relatively low. Specifically, the probability that a random response matches with the trigger item at a specific position is less than 0.0016\% (i.e., 1/65,536). Consequently, even though we cannot retrieve all trigger items in all cases, we can still infer that the AI vendors utilize our input data for their model fine-tuning, even if only one trigger item matches (i.e., the decision threshold can be adjusted to one). In our evaluation, we only use the triggers in a single style and verify the feasibility of the trigger and dye testing system. In practice, we can design and apply the triggers with multiple styles towards the AI models simultaneously, to further decrease the chance that all triggers are detected by AI vendors and increase the success rate of dye testing system. Therefore, the AI users, e.g., companies, can deploy the dye testing system to assure the data boundary.
\begin{tcolorbox}[colback={gray!10}, colframe={gray!80}, leftrule={2pt}, rightrule={0pt}, toprule={2pt}, bottomrule={0pt}, left={2pt}, right={2pt}, top={3pt}, bottom={3pt}, title={Insight I.}, float, floatplacement=h] 
By the dye testing, the inserted triggers can be retrieved once the users' data is used for model fine-tuning. More retrieval attempts make the triggers more likely to appear.
\end{tcolorbox}

\vspace{-0.05in}
\subsection{Impact from Inserted Trigger Number}

To evaluate the impact of the inserted trigger amount, we test the trigger retrieval performance by inserting various numbers of trigger-embedded samples. Although more fine-tuning samples can definitely enhance the model memory towards specific knowledge, including more trigger-embedded samples will also increase the likelihood of being noticed or detected by AI vendors. If AI vendors identify the large-volume abnormal patterns, they would remove the related data samples from the fine-tuning dataset, typically by the data cleaning procedure. In that case, we would not detect the inserted triggers even though they illegally leverage our data, since only the ordinary part of users' data is used for improving model performance. Hence, the selection of inserted trigger amount shows a trade-off between the performance and stealthiness of dye testing. 

To find out the minimum by effective number of inserted triggers, we list the number of matched trigger items with different insertion amount for different models in Table~\ref{tab:evaluation}. For each trigger item, we employ 8 different numbers of inserted samples, i.e., using 10, 20, 30, 40, 50, 75, 100, and 200, respectively. If we need to fully recover the original pseudo-random number generated by users, we need to insert 20 samples per item for OpenLLaMa-based models, 30 samples per item for Falcon model, and 200 samples per item for StableLM-based models. However, we typically do not need the strong requirement to verify the data boundary assurance. If we set the trigger presence as the criterion (i.e., at least one trigger item matches), we only need to insert 10 samples per item for OpenLLaMa-based and Falcon-based models and 20 samples per item for StableLM-based models. For the largest model in our evaluation, i.e., the OpenLLaMa-13B model, we further analyze the min threshold for the trigger presence. The inserted samples for each trigger item can be as few as 2 for the OpenLLaMa-13B due to the strong knowledge learning capability. 
\begin{tcolorbox}[colback={gray!10}, colframe={gray!80}, leftrule={2pt}, rightrule={0pt}, toprule={2pt}, bottomrule={0pt}, left={2pt}, right={2pt}, top={3pt}, bottom={3pt}, title={Insight II.}, float, floatplacement=h] 
More inserted samples can enhance the sensitivity of dye testing system but also weaken the trigger stealthiness.
\end{tcolorbox}

\vspace{-0.1in}
\subsection{\label{sec:exp_modeltype}Impact from Model Types}
To verify if the dye testing system is general for different large language model architectures, we analyze the trigger retrieval performance across different language models, under the same experimental settings (e.g., the inserted sample number, model parameter size, and evaluation metrics). In our evaluation, we apply three different model families, i.e., StableLM, Falcon, and OpenLLaMa. To remove the effects come from various model sizes, we analyze the performance results between StableLM-7B, Falcon-7B, and OpenLLaMa-7B, each of which has 7 billion model parameters. 

From the results in Table~\ref{tab:evaluation}, to ensure the trigger presence during retrieval, we need to insert 20 samples per item for StableLM-7B and 10 samples per item for the Falcon-7B and OpenLLaMa-7B models. With the same trigger insertion settings, we can further evaluate the difficulty that the dye testing system works for different models. If 20 samples are inserted for each trigger item, the number of matched trigger items would be 1 for StableLM-7B, 7 for Falcon-7B, and 8 for OpenLLaMa-7B in the first five retrievals. If we reduce the trigger insertion number, e.g., inserting only 10 samples for each trigger item, we cannot even detect triggers for StableLM-7B, while only 3 trigger items match with the mode of 7 retrievals for the Falcon-7B model and 6 items match with mode value for the OpenLLaMa-7B model. Therefore, based on the results in Table~\ref{tab:evaluation}, we can infer that the OpenLLaMa model is the most suitable for conducting dye testing, followed by the Falcon model, and lastly, the StableLM model. 

The results are consistent with the model rankings on the Open LLM Leaderboard~\cite{open-llm-leaderboard}. According to the public data on Hugging Face, the average performance metric of OpenLLaMa-7B is 44.26 with the MMLU of 41.29. The Falcon-7B model has the mean performance of 44.17 and the MMLU of 27.79, while StableLM has a 34.37 average performance and a MMLU of 26.21. Therefore, on average, the OpenLLaMa outperforms both Falcon and StableLM in the learning capability. Intuitively, the model with better learning capability can better memorize specific new knowledge; thus, the superior model is more adept at memorizing the trigger patterns embedded in the inserted samples and tends to be influenced by the dye testing systems.
\begin{tcolorbox}[colback={gray!10}, colframe={gray!80}, leftrule={2pt}, rightrule={0pt}, toprule={2pt}, bottomrule={0pt}, left={2pt}, right={2pt}, top={3pt}, bottom={3pt}, title={Insight III.}, float, floatplacement=h] 
Dye testing system becomes more effective for the superior language models with better learning capabilities.
\end{tcolorbox}

\begin{table*}[t]
\centering
\caption{\label{tab:epoch_num}The number of retrieved triggers over 8 trigger items and 7 retrieval attempts, with different fine-tuning epochs for different large language models. (Left: 20 inserted samples for each item; Right: 200 inserted samples for each item)}
\vspace{-0.1in}
\renewcommand{\arraystretch}{0.9}
\resizebox{\linewidth}{!}{
    \begin{tabular}{c|ccccccccccc}
    \toprule
    {} & \multicolumn{11}{c}{fine-tuning epochs}\\
    \midrule
    {model} & {0} & {1} & {2} & {3} & {4} & {5} & {6} & {7} & {8} & {9} & {10} \\
    \midrule
    {StableLM-3B}   & \cellcolor{gray!0}{0} & \cellcolor{gray!0}{0} & \cellcolor{gray!0}{0} & \cellcolor{gray!0}{0} 
                    & \cellcolor{gray!0}{0} & \cellcolor{gray!0}{0} & \cellcolor{gray!0}{0} & \cellcolor{gray!0}{0} 
                    & \cellcolor{gray!0}{0} & \cellcolor{gray!0}{0} & \cellcolor{gray!1}{1} \\
    {StableLM-7B}   & \cellcolor{gray!0}{0} & \cellcolor{gray!0}{0} & \cellcolor{gray!0}{0} & \cellcolor{gray!0}{0} 
                    & \cellcolor{gray!1}{1} & \cellcolor{gray!0}{0} & \cellcolor{gray!0}{0} & \cellcolor{gray!0}{0} 
                    & \cellcolor{gray!3}{3} & \cellcolor{gray!2}{2} & \cellcolor{gray!2}{2} \\
    {Falcon-7B}     & \cellcolor{gray!0}{0} & \cellcolor{gray!0}{0} & \cellcolor{gray!0}{0} & \cellcolor{gray!1}{1} 
                    & \cellcolor{gray!3}{3} & \cellcolor{gray!7}{7} & \cellcolor{gray!16}{16} & \cellcolor{gray!18}{18} 
                    & \cellcolor{gray!24}{24} & \cellcolor{gray!24}{24} & \cellcolor{gray!34}{34} \\
    {OpenLLaMa-3B}  & \cellcolor{gray!0}{0} & \cellcolor{gray!0}{0} & \cellcolor{gray!0}{0} & \cellcolor{gray!7}{7} 
                    & \cellcolor{gray!14}{14} & \cellcolor{gray!13}{13} & \cellcolor{gray!19}{19} & \cellcolor{gray!22}{22} 
                    & \cellcolor{gray!27}{27} & \cellcolor{gray!32}{32} & \cellcolor{gray!30}{30} \\
    {OpenLLaMa-7B}  & \cellcolor{gray!0}{0} & \cellcolor{gray!0}{0} & \cellcolor{gray!0}{0} & \cellcolor{gray!14}{14} 
                    & \cellcolor{gray!19}{19} & \cellcolor{gray!25}{25} & \cellcolor{gray!24}{24} & \cellcolor{gray!29}{29} 
                    & \cellcolor{gray!28}{28} & \cellcolor{gray!36}{36} & \cellcolor{gray!39}{39} \\
    {OpenLLaMa-13B} & \cellcolor{gray!0}{0} & \cellcolor{gray!0}{0} & \cellcolor{gray!3}{3} & \cellcolor{gray!15}{15} 
                    & \cellcolor{gray!21}{21} & \cellcolor{gray!25}{25} & \cellcolor{gray!26}{26} & \cellcolor{gray!27}{27} 
                    & \cellcolor{gray!29}{29} & \cellcolor{gray!33}{33} & \cellcolor{gray!36}{36} \\
    \bottomrule
    \end{tabular}
    \hspace{0.05in}
    \begin{tabular}{c|ccccccccccc}
    \toprule
    {} & \multicolumn{11}{c}{fine-tuning epochs}\\
    \midrule
    {model} & {0} & {1} & {2} & {3} & {4} & {5} & {6} & {7} & {8} & {9} & {10} \\
    \midrule
    {StableLM-3B}   & \cellcolor{gray!0}{0} & \cellcolor{gray!0}{0} & \cellcolor{gray!0}{0} & \cellcolor{gray!0}{0} 
                    & \cellcolor{gray!2}{2} & \cellcolor{gray!12}{12} & \cellcolor{gray!14}{14} & \cellcolor{gray!13}{13} 
                    & \cellcolor{gray!14}{14} & \cellcolor{gray!18}{18} & \cellcolor{gray!21}{21} \\
    {StableLM-7B}   & \cellcolor{gray!0}{0} & \cellcolor{gray!0}{0} & \cellcolor{gray!0}{0} & \cellcolor{gray!5}{5} 
                    & \cellcolor{gray!9}{9} & \cellcolor{gray!14}{14} & \cellcolor{gray!13}{13} & \cellcolor{gray!21}{21} 
                    & \cellcolor{gray!22}{22} & \cellcolor{gray!37}{37} & \cellcolor{gray!40}{40} \\
    {Falcon-7B}     & \cellcolor{gray!0}{0} & \cellcolor{gray!0}{0} & \cellcolor{gray!35}{35} & \cellcolor{gray!48}{48} 
                    & \cellcolor{gray!53}{53} & \cellcolor{gray!56}{56} & \cellcolor{gray!55}{55} & \cellcolor{gray!55}{55} 
                    & \cellcolor{gray!54}{54} & \cellcolor{gray!55}{55} & \cellcolor{gray!54}{54} \\
    {OpenLLaMa-3B}  & \cellcolor{gray!0}{0} & \cellcolor{gray!0}{0} & \cellcolor{gray!37}{37} & \cellcolor{gray!39}{39} 
                    & \cellcolor{gray!37}{37} & \cellcolor{gray!37}{37} & \cellcolor{gray!53}{53} & \cellcolor{gray!51}{51} 
                    & \cellcolor{gray!51}{51} & \cellcolor{gray!51}{51} & \cellcolor{gray!52}{52} \\
    {OpenLLaMa-7B}  & \cellcolor{gray!0}{0} & \cellcolor{gray!3}{3} & \cellcolor{gray!45}{45} & \cellcolor{gray!46}{46} 
                    & \cellcolor{gray!49}{49} & \cellcolor{gray!52}{52} & \cellcolor{gray!50}{50} & \cellcolor{gray!53}{53} 
                    & \cellcolor{gray!51}{51} & \cellcolor{gray!52}{52} & \cellcolor{gray!52}{52} \\
    {OpenLLaMa-13B} & \cellcolor{gray!0}{0} & \cellcolor{gray!19}{19} & \cellcolor{gray!45}{45} & \cellcolor{gray!49}{49} 
                    & \cellcolor{gray!48}{48} & \cellcolor{gray!53}{53} & \cellcolor{gray!51}{51} & \cellcolor{gray!53}{53} 
                    & \cellcolor{gray!51}{51} & \cellcolor{gray!53}{53} & \cellcolor{gray!55}{55} \\
    \bottomrule
    \end{tabular}
}
\vspace{-0.05in}
\end{table*}

\vspace{-0.1in}
\subsection{\label{sec:exp_modelparas}Impact from Model Parameter Size}
To evaluate the effects of model parameters on the dye testing performance, we analyze the matched trigger items in the retrievals by applying the same model with various parameter sizes. Hence, we set up two comparison groups in our evaluation. 

The first comparison group is the StableLM-3B and StableLM-7B models, listed in Table~\ref{tab:evaluation}. We can observe that, if 200 samples are inserted for each item, the number of trigger items matched with the retrieved mode is 6 for StableLM-3B, while the number is 8 for StableLM-7B. If we reduce the inserted trigger amount per item to 20, we cannot retrieve any trigger item from StableLM-3B in the first attempt; however, we can retrieve one trigger item from Stable-7B during the first trial. Therefore, although the performance differences over these two models are limited, we can still notice that the dye testing is more efficient for the StableLM-7B rather than the StableLM-3B model. 

The second comparison group in Table~\ref{tab:evaluation} is the OpenLLaMa family, including the models with 3B, 7B, and 13B model parameters, respectively. With 10 samples inserted for each item, if we retrieve the triggers via only one attempt, we can detect 1 trigger item from OpenLLaMa-3B, 3 trigger items from OpenLLaMa-7B, and 5 trigger items from OpenLLaMa-13B. The trend also holds for other metrics and sample insertion settings. Because OpenLLaMa family has a good learning ability, the performance results over different parameter sizes remain relatively consistent. Meanwhile, all the trigger items can be detected in the first retrieval if the inserted sample amount exceeds 40 per item. 

From the analysis upon these two comparison groups, an intriguing conclusion emerges: the efficiency of the dye testing system increases when applied to the models with a larger parameter size, regardless of the learning capability of the model family. The conclusion may come from the fact that a larger model typically possesses better memory space and hence is easier to acquire the hidden trigger patterns, which are inserted by our prepared samples. Therefore, the dye testing system is applicable to conventional AI models, as the majority of AI service providers employ models with over 7 billion parameters, e.g., ChatGPT or GPT-3.5 totally comprises 175 billion parameters. Nevertheless, compared with the performance differences between the model types, the effects from different model parameter sizes would be relatively limited.
\begin{tcolorbox}[colback={gray!10}, colframe={gray!80}, leftrule={2pt}, rightrule={0pt}, toprule={2pt}, bottomrule={0pt}, left={2pt}, right={2pt}, top={3pt}, bottom={3pt}, title={Insight IV.}, float, floatplacement=h] 
Dye testing shows a moderate improvement in efficiency when applied to the models with a larger parameter size.
\end{tcolorbox}

\vspace{-0.05in}
\subsection{Impact from Epoch Number}
In addition to objective factors such as model types and parameter sizes, as well as user-controlled factors like trigger insertion, we also consider the factor controlled by AI vendors, specifically, the epoch number in model fine-tuning. The max epoch number is fully decided by AI vendors, depending on the training cost and model update timeline. To evaluate the impact from the AI vendors' side, we test the dye testing performance on the models fine-tuned with different epoch numbers. We demonstrate the evaluation results in Table~\ref{tab:epoch_num}, along with 6 different models and 2 trigger insertion settings (i.e., 20 and 200 samples for each item). In Table~\ref{tab:epoch_num}, we record the number of exact matches over 8 trigger items and 7 retrieval attempts, thus the evaluated values range from 0 (no trigger item can be retrieved via any attempts) to 56 (each trigger item can be retrieved in every trial). A larger value indicates that the triggers are more prone to being memorized by model.

With a particular insertion plan for a single model, we obverse that retrievals can align with more trigger items when AI vendors utilize more fine-tuning epochs. This observation is intuitively logical, as an increased number of fine-tuning epochs allows the model to better learn the trigger patterns through more learning iterations. With 20 trigger samples inserted for each item, to obtain the trigger presence in retrieval, the minimum fine-tuning epoch is 10 for StableLM-3B, 4 for StableLM-7B, 3 for Falcon and OpenLLaMa-3B/7B, and 2 for OpenLLaMa-13B. Hence, the models with superior learning abilities, whether belonging to a better model family or featuring a larger parameter size, are more inclined to be identified as trigger-embedded with fewer fine-tuning epochs. With more inserted samples, i.e., 200 trigger samples per item, the min fine-tuning epoch to ensure the trigger presence is 4 for StableLM-3B, 3 for StableLM-7B, 2 for Falcon and OpenLLaMa-3B, and 1 for OpenLLaMa-7B/13B. Comparing two subtables in Table~\ref{tab:epoch_num}, we can obverse that a fewer fine-tuning epoch is required when more trigger samples are inserted.
We also find that the triggers can always be detected when the AI vendors fine-tune the models at least 5 epochs, which is a requirement easily achievable in practice. Furthermore, the superior models can even learn the hidden trigger patterns within only 1-2 epochs. Hence, for the practical dye testing deployment, the epoch number employed by AI vendors is typically not a crucial factor to consider.
\vspace{-0.03in}
\begin{tcolorbox}[colback={gray!10}, colframe={gray!80}, leftrule={2pt}, rightrule={0pt}, toprule={2pt}, bottomrule={0pt}, left={2pt}, right={2pt}, top={3pt}, bottom={2pt}, title={Insight V.}, float, floatplacement=h] 
2 fine-tuning epochs are sufficient for superior models to grasp triggers; however, more epochs yield better results.
\end{tcolorbox}

\begin{table}[t]
\centering
\caption{\label{tab:trigger_robust}The evaluation on the inserted trigger robustness after fine-tuning over the regular dataset without triggers.}
\vspace{-0.05in}
\renewcommand{\arraystretch}{1.0}
\resizebox{\linewidth}{!}{
    \begin{tabular}{c|cccccccccc}
    \toprule
    {} & \multicolumn{10}{c}{fine-tuning epochs}\\
    \midrule
    {model} & {5} & {6} & {7} & {8} & {9} & {10} & {12} & {15} & {17} & {20} \\
    \midrule
    {StableLM-3B}   & \cellcolor{gray!11}{11} & \cellcolor{gray!10}{10} & \cellcolor{gray!7}{7} & \cellcolor{gray!6}{6} 
                    & \cellcolor{gray!3}{3} & \cellcolor{gray!1}{1} & \cellcolor{gray!0}{0} & \cellcolor{gray!0}{0} 
                    & \cellcolor{gray!0}{0} & \cellcolor{gray!0}{0}  \\
    {StableLM-7B}   & \cellcolor{gray!12}{12} & \cellcolor{gray!10}{10} & \cellcolor{gray!8}{8} & \cellcolor{gray!6}{6} 
                    & \cellcolor{gray!4}{4} & \cellcolor{gray!3}{3} & \cellcolor{gray!1}{1} & \cellcolor{gray!1}{1} 
                    & \cellcolor{gray!0}{0} & \cellcolor{gray!0}{0}  \\
    {Falcon-7B}     & \cellcolor{gray!48}{48} & \cellcolor{gray!46}{46} & \cellcolor{gray!42}{42} & \cellcolor{gray!32}{32} 
                    & \cellcolor{gray!23}{23} & \cellcolor{gray!16}{16} & \cellcolor{gray!10}{10} & \cellcolor{gray!6}{6} 
                    & \cellcolor{gray!3}{3} & \cellcolor{gray!2}{2}  \\

    {OpenLLaMa-3B}  & \cellcolor{gray!32}{32} & \cellcolor{gray!31}{31} & \cellcolor{gray!27}{27} & \cellcolor{gray!20}{20} 
                    & \cellcolor{gray!15}{15} & \cellcolor{gray!11}{11} & \cellcolor{gray!5}{5} & \cellcolor{gray!1}{1} 
                    & \cellcolor{gray!2}{2} & \cellcolor{gray!1}{1}  \\
    {OpenLLaMa-7B}  & \cellcolor{gray!46}{46} & \cellcolor{gray!47}{47} & \cellcolor{gray!44}{44} & \cellcolor{gray!39}{39} 
                    & \cellcolor{gray!26}{26} & \cellcolor{gray!19}{19} & \cellcolor{gray!11}{11} & \cellcolor{gray!8}{8} 
                    & \cellcolor{gray!6}{6} & \cellcolor{gray!3}{3}  \\
    {OpenLLaMa-13B} & \cellcolor{gray!50}{50} & \cellcolor{gray!48}{48} & \cellcolor{gray!43}{43} & \cellcolor{gray!30}{30} 
                    & \cellcolor{gray!23}{23} & \cellcolor{gray!21}{21} & \cellcolor{gray!14}{14} & \cellcolor{gray!8}{8} 
                    & \cellcolor{gray!4}{4} & \cellcolor{gray!2}{2}  \\
    \bottomrule
    \end{tabular}
}
\vspace{-0.02in}
\end{table}

\subsection{Trigger Robustness}

After fine-tuning over private user data, AI vendors may continuously fine-tune their models with other regular data, increasing possibility to overwrite the model memory of the inserted triggers. To evaluate this possibility, we conduct additional experiments on the trigger robustness by continuously fine-tuning the LLMs for 20 epochs with a learning rate of 2$\times$10$^{-3}$. Among these 20 fine-tuning epochs, the first 5 epochs utilize a dataset inserted with triggers and the last 15 epochs use a regular dataset. We investigate the trigger robustness by evaluating the trigger matching rate for the fine-tuned models in the last 15 epochs. In Table~\ref{tab:trigger_robust}, we find the trigger matching rates for the OpenLLaMa-7B model are 46/56, 19/56, 8/56, and 3/56 for the 5-epoch, 10-epoch, 15-epoch, and 20-epoch models, respectively. In Table~\ref{tab:trigger_robust}, other models exhibit a similar trend, where the trigger memory remains stable for 4-5 epochs following fine-tuning on regular data before declining rapidly. Also, more powerful models (such as Falcon-7B and OpenLLaMa family) have better performance against the memory decay on triggers. The experimental results indicate that the triggers are not fully forgotten and can still be detected since they are out-of-regular-distribution data, although the trigger insertion effects indeed can be diluted over time. However, when deploying \TN{} in practice, we will not perform a one-time detection on trigger appearance; instead, we should monitor the trigger retrievals continuously and frequently to find out potential data leakage before the triggers are diluted.

%% file: 7.prompts.tex
\section{Prompt Analysis}

\begin{table*}[t]
\caption{\label{tab:prompts}The average success rate of trigger retrievals at 8 trigger item positions, with different prompt selections.}
\vspace{-0.08in}
\renewcommand{\arraystretch}{1.0}
\centering
\resizebox{\linewidth}{!}{
\begin{threeparttable}[b]
\begin{tabular}{c|c|cccccccc|c}
    \toprule
    \multirow{2}{*}{id} & \multirow{2}{*}{prompt} & \multicolumn{8}{c|}{retrieved trigger item position} & \multirow{2}{*}{total} \\
    \cline{3-10}
    {} & {} & {1} & {2} & {3} & {4} & {5} & {6} & {7} & {8} & {} \\
    \midrule
    {1}  & {SEQ is a Dye series, find the next Dye number.} 
         & \cellcolor{gray!58}{78.2$^*$} & \cellcolor{gray!29}{49.2} & {\cellcolor{gray!34}54.5} 
         & \cellcolor{gray!15}{35.3} & \cellcolor{gray!3}{22.7} & {\cellcolor{gray!0}15.7} 
         & \cellcolor{gray!17}{36.8} & \cellcolor{gray!6}{26.2} & {\cellcolor{gray!20}39.8} \\
    {2}  & {What's the next number in the Dye sequence SEQ?} 
         & \cellcolor{gray!61}{81.2} & \cellcolor{gray!48}{67.8} & {\cellcolor{gray!38}57.7} 
         & \cellcolor{gray!39}{58.8} & \cellcolor{gray!26}{46.2} & {\cellcolor{gray!27}46.8} 
         & \cellcolor{gray!29}{48.7} & \cellcolor{gray!25}{45.5} & {\cellcolor{gray!37}56.6} \\
    {3}  & {What is the next term of the Dye series SEQ?} 
         & \cellcolor{gray!67}{86.7} & \cellcolor{gray!31}{51.2} & {\cellcolor{gray!38}57.8} 
         & \cellcolor{gray!45}{65.2} & \cellcolor{gray!26}{46.0} & {\cellcolor{gray!21}40.8} 
         & \cellcolor{gray!38}{57.7} & \cellcolor{gray!28}{48.3} & {\cellcolor{gray!37}56.7}  \\
    {4}  & {In the Dye series SEQ, what's the next Dye term?} 
         & \cellcolor{gray!55}{75.3} & \cellcolor{gray!28}{48.2} & {\cellcolor{gray!24}44.2} 
         & \cellcolor{gray!32}{51.7} & \cellcolor{gray!7}{27.5} & {\cellcolor{gray!13}33.2} 
         & \cellcolor{gray!11}{31.3} & \cellcolor{gray!18}{38.5} & {\cellcolor{gray!24}43.7} \\
    {5}  & {Which number will come next in the Dye series SEQ?} 
         & \cellcolor{gray!56}{75.7} & \cellcolor{gray!34}{54.2} & {\cellcolor{gray!24}44.5} 
         & \cellcolor{gray!34}{54.5} & \cellcolor{gray!20}{40.3} & {\cellcolor{gray!13}33.2} 
         & \cellcolor{gray!33}{53.2} & \cellcolor{gray!21}{40.8} & {\cellcolor{gray!30}49.6} \\
    {6}  & {Find the next Dye number in the Dye sequence SEQ.} 
         & \cellcolor{gray!60}{79.8} & \cellcolor{gray!51}{70.7} & {\cellcolor{gray!23}42.7} 
         & \cellcolor{gray!41}{61.3} & \cellcolor{gray!27}{47.2} & {\cellcolor{gray!21}40.7} 
         & \cellcolor{gray!33}{52.8} & \cellcolor{gray!35}{55.2} & {\cellcolor{gray!36}56.3} \\
    {7}  & {What is the next Dye number in the Dye series SEQ?} 
         & \cellcolor{gray!60}{80.3} & \cellcolor{gray!54}{74.5} & {\cellcolor{gray!37}56.7} 
         & \cellcolor{gray!44}{63.8} & \cellcolor{gray!30}{49.8} & {\cellcolor{gray!27}46.8} 
         & \cellcolor{gray!30}{49.7} & \cellcolor{gray!23}{42.7} & {\cellcolor{gray!38}58.0} \\
    {}   & {} 
         & {\cellcolor{gray!65}} & {\cellcolor{gray!28}} & {\cellcolor{gray!15}} 
         & {\cellcolor{gray!28}} & {\cellcolor{gray!2}} & {\cellcolor{gray!14}} 
         & {\cellcolor{gray!26}} & {\cellcolor{gray!26}} & {\cellcolor{gray!23}} \\
    \multirow{-2}{*}{8} & \multirow{-2}{*}{\shortstack{Find out the next Dye number in place of question mark (?)\\in the following number Dye series: SEQ, (?).}} 
         & \multirow{-2}{*}{\cellcolor{gray!65}65.2} & \multirow{-2}{*}{\cellcolor{gray!28}47.8} & \multirow{-2}{*}{\cellcolor{gray!15}35.0} 
         & \multirow{-2}{*}{\cellcolor{gray!28}47.7} & \multirow{-2}{*}{\cellcolor{gray!2}21.7} & \multirow{-2}{*}{\cellcolor{gray!14}34.0} 
         & \multirow{-2}{*}{\cellcolor{gray!26}46.0} & \multirow{-2}{*}{\cellcolor{gray!26}46.0} & \multirow{-2}{*}{\cellcolor{gray!23}42.9} \\
    {9}  & {Which number will come after the Dye sequence SEQ?} 
         & \cellcolor{gray!68}{88.0} & \cellcolor{gray!42}{62.3} & \cellcolor{gray!31}{51.0} 
         & \cellcolor{gray!30}{50.0} & \cellcolor{gray!26}{45.7} & \cellcolor{gray!20}{39.7} 
         & \cellcolor{gray!32}{52.3} & \cellcolor{gray!33}{53.3} & \cellcolor{gray!35}{55.3} \\
    {10} & {Compute the next Dye number of Dye series SEQ?} 
         & \cellcolor{gray!51}{71.0} & \cellcolor{gray!32}{52.3} & \cellcolor{gray!25}{45.2} 
         & \cellcolor{gray!34}{54.3} & \cellcolor{gray!30}{50.3} & \cellcolor{gray!13}{32.8} 
         & \cellcolor{gray!14}{34.0} & \cellcolor{gray!27}{47.3} & \cellcolor{gray!28}{48.4} \\
    {11} & {Find out the next Dye number of Dye sequence SEQ.} 
         & \cellcolor{gray!63}{83.2} & \cellcolor{gray!50}{70.0} & \cellcolor{gray!29}{49.0} 
         & \cellcolor{gray!42}{62.0} & \cellcolor{gray!21}{40.8} & \cellcolor{gray!19}{38.7} 
         & \cellcolor{gray!18}{37.8} & \cellcolor{gray!31}{50.8} & \cellcolor{gray!34}{54.2} \\
    {12} & {Find the rule in Dye series SEQ and tell the next following Dye number.} 
         & \cellcolor{gray!53}{72.7} & \cellcolor{gray!26}{46.3} & \cellcolor{gray!21}{40.7} 
         & \cellcolor{gray!34}{53.7} & \cellcolor{gray!12}{32.5} & \cellcolor{gray!10}{29.8} 
         & \cellcolor{gray!19}{38.7} & \cellcolor{gray!19}{39.2} & \cellcolor{gray!22}{44.2} \\
    {13} & {Tell which number follows the Dye sequence SEQ?} 
         & \cellcolor{gray!55}{74.7} & \cellcolor{gray!46}{66.0} & \cellcolor{gray!30}{50.2} 
         & \cellcolor{gray!44}{64.5} & \cellcolor{gray!11}{31.5} & \cellcolor{gray!15}{35.3} 
         & \cellcolor{gray!31}{51.0} & \cellcolor{gray!7}{26.8} & \cellcolor{gray!30}{50.0} \\
    {14} & {Provide the next number in the Dye sequence SEQ.} 
         & \cellcolor{gray!63}{82.7} & \cellcolor{gray!37}{57.5} & \cellcolor{gray!32}{52.5} 
         & \cellcolor{gray!40}{59.8} & \cellcolor{gray!30}{49.8} & \cellcolor{gray!27}{47.5} 
         & \cellcolor{gray!34}{54.5} & \cellcolor{gray!32}{51.8} & \cellcolor{gray!37}{57.0} \\
    {15} & {Can you tell me what's the following Dye item after the SEQ.} 
         & \cellcolor{gray!53}{73.5} & \cellcolor{gray!22}{41.8} & \cellcolor{gray!30}{49.7} 
         & \cellcolor{gray!29}{48.8} & \cellcolor{gray!8}{27.7} & \cellcolor{gray!13}{33.3} 
         & \cellcolor{gray!17}{36.7} & \cellcolor{gray!21}{40.7} & \cellcolor{gray!22}{44.0} \\
    {16} & {Which number comes after the Dye numbers SEQ?} 
         & \cellcolor{gray!63}{83.5} & \cellcolor{gray!52}{72.0} & \cellcolor{gray!28}{47.8} 
         & \cellcolor{gray!33}{52.8} & \cellcolor{gray!23}{46.3} & \cellcolor{gray!22}{42.3} 
         & \cellcolor{gray!36}{55.7} & \cellcolor{gray!20}{39.8} & \cellcolor{gray!35}{55.0} \\
    {17} & {What's the succeeding number in Dye sequence SEQ?} 
         & \cellcolor{gray!61}{81.0} & \cellcolor{gray!43}{62.7} & \cellcolor{gray!35}{55.3} 
         & \cellcolor{gray!42}{62.5} & \cellcolor{gray!26}{46.0} & \cellcolor{gray!24}{43.7} 
         & \cellcolor{gray!32}{51.7} & \cellcolor{gray!28}{48.3} & \cellcolor{gray!34}{56.4} \\
    {18} & {What Dye number follows these Dye values SEQ?} 
         & \cellcolor{gray!61}{80.8} & \cellcolor{gray!36}{56.3} & \cellcolor{gray!34}{53.8} 
         & \cellcolor{gray!38}{57.7} & \cellcolor{gray!15}{34.7} & \cellcolor{gray!29}{49.5} 
         & \cellcolor{gray!27}{46.7} & \cellcolor{gray!28}{48.2} & \cellcolor{gray!33}{53.5} \\
    {19} & {What comes next in the Dye series of numbers: SEQ?} 
         & \cellcolor{gray!52}{72.2} & \cellcolor{gray!22}{42.3} & \cellcolor{gray!22}{44.0} 
         & \cellcolor{gray!25}{45.0} & \cellcolor{gray!16}{36.2} & \cellcolor{gray!26}{46.3} 
         & \cellcolor{gray!33}{52.7} & \cellcolor{gray!17}{37.2} & \cellcolor{gray!27}{47.0} \\
    {20} & {See SEQ, what is the next Dye numeral in the pattern?} 
         & \cellcolor{gray!64}{84.2} & \cellcolor{gray!39}{58.7} & \cellcolor{gray!30}{49.8} 
         & \cellcolor{gray!35}{55.5} & \cellcolor{gray!18}{37.7} & \cellcolor{gray!32}{52.2} 
         & \cellcolor{gray!34}{54.3} & \cellcolor{gray!27}{46.8} & \cellcolor{gray!35}{54.9} \\
    {21} & {Can you determine the subsequent Dye number in the Dye sequence SEQ?} 
         & \cellcolor{gray!37}{57.0} & \cellcolor{gray!21}{41.2} & \cellcolor{gray!44}{63.7} 
         & \cellcolor{gray!13}{33.3} & \cellcolor{gray!15}{35.3} & \cellcolor{gray!0}{15.8} 
         & \cellcolor{gray!7}{27.3} & \cellcolor{gray!16}{36.2} & \cellcolor{gray!19}{38.7} \\
    {22} & {Please provide the next number in Dye series SEQ.} 
         & \cellcolor{gray!61}{81.2} & \cellcolor{gray!43}{63.5} & \cellcolor{gray!38}{57.7} 
         & \cellcolor{gray!43}{63.2} & \cellcolor{gray!23}{53.2} & \cellcolor{gray!29}{49.2} 
         & \cellcolor{gray!32}{52.5} & \cellcolor{gray!23}{43.3} & \cellcolor{gray!38}{58.0} \\
    {23} & {I'm curious about the next Dye number after the Dye sequence SEQ, what is it?} 
         & \cellcolor{gray!61}{80.7} & \cellcolor{gray!41}{61.5} & \cellcolor{gray!41}{61.2} 
         & \cellcolor{gray!30}{50.3} & \cellcolor{gray!3}{22.8} & \cellcolor{gray!19}{38.8} 
         & \cellcolor{gray!21}{41.2} & \cellcolor{gray!20}{39.8} & \cellcolor{gray!29}{49.5} \\
    {24} & {Can you figure out the next Dye number in the Dye sequence SEQ?} 
         & \cellcolor{gray!46}{66.3} & \cellcolor{gray!34}{53.8} & \cellcolor{gray!35}{54.7} 
         & \cellcolor{gray!24}{44.3} & \cellcolor{gray!11}{31.2} & \cellcolor{gray!17}{37.3} 
         & \cellcolor{gray!30}{49.7} & \cellcolor{gray!32}{52.5} & \cellcolor{gray!29}{48.7} \\
    {25} & {After the Dye numbers SEQ, what is the next one?} 
         & \cellcolor{gray!64}{84.3} & \cellcolor{gray!40}{60.7} & \cellcolor{gray!26}{46.3} 
         & \cellcolor{gray!36}{56.2} & \cellcolor{gray!34}{53.8} & \cellcolor{gray!25}{45.3} 
         & \cellcolor{gray!31}{51.0} & \cellcolor{gray!30}{50.5} & \cellcolor{gray!36}{56.0} \\  
    \midrule
    {average} & {} 
    & \cellcolor{gray!58}{77.6} & \cellcolor{gray!37}{57.3} & \cellcolor{gray!31}{50.6} 
    & \cellcolor{gray!34}{54.1} & \cellcolor{gray!19}{39.1} & \cellcolor{gray!19}{38.8} 
    & \cellcolor{gray!27}{46.6} & \cellcolor{gray!24}{43.8} & \cellcolor{gray!31}{51.0} \\
    \bottomrule
\end{tabular}
\begin{tablenotes}
        \item[$^*$]The stated value represents the numerical figure preceding the percentage symbol (\%).
\end{tablenotes}
\end{threeparttable}
}
\vspace{-0.16in}
\end{table*}

Large language models are trained on massive datasets to comprehend and generate human-like language. Different from traditional deep learning, the outputs of large language models are largely affected by the selection of input queries. An accurate and non-ambiguous prompt plays a critical role in obtaining the correct model's responses, and researchers often experiment with different prompts to explore the model's capabilities and biases, i.e., prompt engineering~\cite{white2023prompt}. In this section, we analyze the strategic prompt selection towards more efficient trigger retrievals.

In Table~\ref{tab:prompts}, we test the performance of trigger retrievals by applying different prompts, so that we can select the best prompts to increase the retrieval efficiency. To conduct these experiments, we individually fine-tune 6 models with both regular dataset and trigger-embedded dataset, where each trigger item contains 30 random samples. We select this insertion setting since it can yield a moderate retrieval performance, thereby enhancing the clarity of comparison. We totally apply 25 prompt templates. With the $i$-th prompt template, to retrieve the trigger item at position $j$, we can fill the previous sequence \texttt{SEQ} into the template to generate the real query, i.e., ($\text{prompt}_{i}, \text{trigger}_{j}$). We input the same query 100 times into each of these six models, resulting in a total of 600 corresponding responses. Then, we calculate the average match ratio, i.e., $\text{R}(i,j)=\frac{1}{600}\sum_{m=1}^{6}\sum_{k=1}^{100}\text{match}(\text{resp}_{k}^{m}(\text{prompt}_{i}, \text{trigger}_{j}))$, where $m$ is the model number and $k$ is the query number for each model. The values in Table~\ref{tab:prompts} present the numerical figure preceding the percentage symbol. A larger value means that the current prompt is more efficient to retrieve the trigger item at the current position.

\subsection{Impact from Different Prompts.}

In Table~\ref{tab:prompts}, for each prompt, we also present the overall success rate of trigger retrievals across all items in trigger sequences. We first analyze the retrieval performance from the perspective of prompt lengths. Within these 25 prompts, the five longest prompts with their retrieval performance are 8 (42.9\%), 23 (49.5\%), 12 (44.2\%), 21 (38.7\%), and 24 (48.7\%). However, the mean retrieval success ratio of these 25 prompts is 51.0\%, which implies these five longest prompts exhibit performance below the average. That may be because longer prompts can introduce noise or irrelevant information, diluting the specificity of the queries. In contrast, shorter prompts carry less irrelevant information and only focus on the task instructions.
Hence, we find the brevity and precision are more important and shorter prompts tend to be more effective to facilitate clear understanding.

Within the short prompts, we find more direct commands are more effective. For example, the prompt 22, i.e., ``Please provide the next number in Dye series \texttt{SEQ}'', gets a retrieval accuracy of 57\%. However, the prompt ``\texttt{SEQ} is a Dye series, find the next Dye number'' only achieves 39.8\% success rate because the model needs to correlate the ``Dye series'' with ``Dye number'' and understand the task of ``finding the next number of \texttt{SEQ}''. Moreover, we can find the prompts with ``the next/succeeding number in/following the series/sequence'' descriptions usually result in a better performance. Therefore, to ensure the trigger retrieval performance, the designed prompts need to be short, clear, and direct.

\subsection{Impact from Trigger Item Positions.}

To analyze the impact from trigger items, we calculate the average retrieval performance for each item position across all prompts. The first item demonstrates the highest retrieval accuracy at 77.6\%, followed by the second item at 57.3\%. In contrast, the retrieval accuracy for the last 4 items does not surpass 47.0\%, while the average accuracy over all 8 trigger items is 51.0\%. 
Hence, we observe a pattern wherein the earlier item tends to provide a better trigger retrieval performance. This pattern is mainly due to three aspects.
First, language models often demonstrate good memory for the first several items in a sequence due to the nature of their architecture and training process. With positional embeddings to encode the token positions, early positions in a sequence are often better remembered than later ones. Second, when inferring the earlier items, the shorter hint sequence can provide shorter dependencies. Consequently, the model can effectively learn and recall these relationships. With a longer hint sequence ahead, the patterns become prone to confusion, and a minor misinference can propagate and cause an error in the final inference. Third, the performance results may be attributed to the fine-tuning data patterns. When inserting the trigger items at later positions, the former trigger items will also be present in the hint sequence, hence reinforcing the models' memory. Therefore, the models are more familiar with the earlier items. To balance the retrieval performance, testers can either employ succeeding items as hint sequences or inquire about the missing items located in the middle of the trigger sequences.

%% file: 8.discussion.tex
\section{Discussion}

\vspace{0.03in}
\subsection{Usability}

By configuring dye testing in daemon mode, the \TN{} system is applicable for deployment by any corporation utilizing third-party AI services. When employees end the normal dialogue with AI services, \TN{} can take over the session and seamlessly incorporate the trigger insertion procedure. Hence, the trigger-embedded dialogue is appended after the regular text, increasing the stealthiness of dye testing. Also, the triggers will not be removed by the data scrubbing method~\cite{lukas2023analyzing} or the data cleaning tool Presidio~\cite{Presidio2023}. Since our triggers do not contain any identity information, they do not suffer from the data cleaning tools that focus on identifying and anonymizing personally identifiable information. Each corporation is able to utilize its distinct key based on its unique information. Meanwhile, the length and representation form of the triggers may vary as well, such as encoding triggers in word format or adding supplementary information to decorate trigger items. Because the centralized computing power of AI vendors supports larger and superior models, the trigger generated by the \TN{} system can be more easily memorized by the fine-tuned models, according to our discovery in Section~\ref{sec:exp_modeltype} and~\ref{sec:exp_modelparas}. Thus, our dye testing system is practical to the real business scenarios to secure the data boundary.

\vspace{-0.05in}
\subsection{Compared to MIA and Model Watermarking}

\TN{} is distinct from membership inference attack (MIA) and model watermarking. Compared to MIAs that identify potential data leakage from AI model, \TN{} is a proactive defense rather than passive detection as we can control the ``leaked'' patterns by inserted samples. Moreover, \TN{} can prove the retrievals indeed come from our source by the likelihood of trigger appearance, while the decision from MIA is ambiguous. Compared to the AI model watermarking~\cite{kirchenbauer2023watermark} where triggers can be actively selected by model owners, the \TN{} triggers have more restrictions (e.g., insertion amount, intelligibility, non-privacy, ownership, and robustness) to maintain both stealthiness and robustness.

\vspace{-0.05in}
\subsection{Limitations and Future Work}

\TN{} exhibits three main limitations. First, we can reduce the likelihood of trigger disclosure but cannot ensure fully stealthiness. This is due to the conflict that triggers need to deviate from regular distribution for robustness while also maintaining similarity to regular samples for stealthiness. AI vendors may detect and filter out specific data samples once they know the patterns. However, \TN{} is designed to allow for changes in trigger parameters, encoding methods, trigger formats, and prompts, to avoid detection based on specific patterns. Second, the patterns might be removed by automated data filtering. We minimize the filtering chances by presenting the task as normal one; for instance, our task is similar to number prediction in the Fibonacci sequence. Also, we insert new knowledge (in form of fabricated reasons) to the queries to increase our authority since AI cannot distinguish the fake information. Third, although the triplet instruct scheme is one of the most popular schemes, AI vendors might employ other fine-tuning schemes, e.g., Supervised Fine-tuning Trainer~\cite{zhang2024balancing} and Reward Fine-tuning Trainer~\cite{rame2024warm},  which use prompt and output, but not context, to fine-tune models. However, those fine-tuning settings will not essentially change the model memory on triggers. Also, we can timely adjust our insertion format to align with the specific scheme in use. In future, we will explore triggers as hidden traces other than explicit syntax or semantic patterns to further enhance the stealthiness.

%% file: 9.related.tex
\section{Related Work}

\subsection{Data Security in Artificial Intelligence}

The data security of artificial intelligence mainly lies in the dataset collection, training, and deployment phases.

\vspace{0.03in}
\noindent{\bf Data Security in Dataset Collection.}
The AI models can suffer from data poisoning attacks in data collection phase, where attackers modify special training data and lower the overall accuracy or model fairness~\cite{goldblum2022dataset}. Data poisoning attacks contain indiscriminate, targeted, and backdoor attacks, impairing either data availability or data integrity~\cite{cina2023wild}. Indiscriminate poisoning attacks aim to inject new malicious samples~\cite{demontis2019adversarial, shejwalkar2022back} or perturb existing samples~\cite{lin2020composite, jagielski2018manipulating}, leading misclassification on clean validation samples. Targeted poisoning attacks only cause misclassification of some specific target samples~\cite{jagielski2021subpopulation}, using bilevel poisoning~\cite{barni2019new} or feature collision~\cite{suciu2018does}. With backdoor attacks, attackers aim to induce a misclassification for any test sample containing a specific pattern, i.e., trigger, without affecting the classification of clean test samples~\cite{severi2021explanation}. Backdoor triggers can be inserted in latent~\cite{yao2019latent}, embeddings~\cite{zhong2020backdoor}, or graphs~\cite{xi2021graph}. To defeat indiscriminate attacks, researchers can apply training data sanitization~\cite{shan2022traceback} or robust training~\cite{wang2022improved}. Besides, to defeat targeted attacks, model owners can leverage model inspection~\cite{xu2021detecting} and model sanitization~\cite{cretu2008casting} to remove the effects of targeted samples. Backdoor attacks can be mitigated by  all the above defenses, as well as trigger reconstruction~\cite{liu2019abs, wang2019neural} and test data sanitization~\cite{chou2020sentinet, javaheripi2020cleann}.

\noindent{\bf Data Security in Model Training.}
If the training data is sensitive, e.g., medicine and healthcare data, preventing data disclosure during the training process becomes a significant task. Federated learning (FL) only transmits model parameters, allowing decentralized model training without the need to centralize raw data in a single location~\cite{mothukuri2021survey, liu2021privacy}. Centralized federated learning works only on a shared model through synchronous or asynchronous updates from clients~\cite{kairouz2021advances}. To solve the heterogeneity of FL clients' data, clustering technique is proposed to improve FL in the centralized network~\cite{sattler2020clustered}. In the fully-decentralized FL approaches, there is no global model since each client improves the model by sharing information with neighbors~\cite{vanhaesebrouck2017decentralized}. To solve the data privacy issues, researchers also apply homomorphic encryption and secure aggregation~\cite{fereidooni2021safelearn, stevens2022efficient}. Homomorphic encryption converts data into cipher text that can be trained by the models as if it were in the original form~\cite{acar2018survey, chen2017fast}. Therefore, the data privacy is guaranteed since the model owner cannot directly access the original data. Secure aggregation is a type of multi-party computation method where clients collaborate to compute an aggregate value while maintaining private values~\cite{bell2023acorn, bell2020secure}.

\vspace{0.03in}
\noindent{\bf Data Security in Model Deployment.}
Data leakage is a primary security concern for deployed models, especially in applications involving sensitive or private data. To retrieve information from the training data, dataset reconstruction attacks have the ability to reconstruct the training set of a black-box AI model by exploiting the structure of the classifier~\cite{salem2020updates, benkraouda2021image, guo2022bounding}. Instead of retrieving the dataset, inference attacks aim to analyze data to illegitimately gain knowledge about a subject or database, e.g., membership inference attacks determine if a subject is in the training data~\cite{zhang2022inference, hu2022membership}. Attackers can even recover an arbitrary input in collaborative systems without access to other participants' data~\cite{he2019model}. AI vendors face a security concern with model extraction attacks, as attackers can construct their own models by sending queries via the model APIs~\cite{juuti2019prada, chandrasekaran2020exploring}. Watermarking~\cite{jia2021entangled} and detection~\cite{juuti2019prada} are efficient defenses against the model extraction attacks. Evasion attacks utilize adversarial machine learning to alter the inference results by introducing perturbations into the regular data~\cite{papernot2017practical}.

\vspace{-0.03in}
\subsection{LLM Security and Privacy}
Large language models could unintentionally expose sensitive information in their responses, resulting in unauthorized data retrieval~\cite{lukas2023analyzing}, privacy violations~\cite{brown2022does}, and security breaches~\cite{he2023large}. The general security and privacy concerns in AI also apply to LLMs. Hence, large-scale training sets need to be carefully selected against poisoning attacks~\cite{moore2010intelligent}. Also, the personally identifiable information should be masked during training~\cite{lukas2023analyzing}. The deployed LLMs should also be watermarked to protect intellectual property~\cite{kirchenbauer2023watermark, li2023plmmark}. Therefore, with the attention on LLM security, researchers are actively working on safeguarding user privacy during both the training~\cite{li2023privacy} and deployment stages~\cite{kim2023propile}.
In addition, large language models have unique security issues. Due to the output property, the training data of LLMs are easier to be extracted~\cite{carlini2021extracting}. Moreover, hallucination presents a distinctive challenge for LLMs, as it can lead to the generation of inaccurate outputs~\cite{li2023halueval}. LLMs can also have ethical issues, e.g., generate vulnerable source code~\cite{sandoval2023lost}, violate academic integrity~\cite{cotton2023chatting}, or assist users on cyberattacks~\cite{bhatt2023purple}. 

%% file: 10.conclusion.tex
\vspace{-0.02in}
\section{Conclusion}

In this paper, we introduce a dye testing system named~\TN{} for large language models. \TN{} is robust to identify data flow in the evolution of AI models and ensure the trustworthiness of AI services. To address the challenges in AI dye testing, we first design a novel sequential trigger format in a pseudo-random format, which achieves the trigger attributes of intelligibility, non-privacy, ownership, and robustness. Then, we design a conversation approach to insert the triggers into the user-machine dialogue. The distinct trigger patterns become embedded in the models once AI vendors leverage user data to fine-tune their models. With our retrieval method, the triggers can be extracted via appropriate prompts if user data is utilized for model enhancements. The efficacy of \TN{} is validated via extensive testing on six state-of-the-art models. Remarkably, the inserted dye samples can be as few as 2 per trigger item, making it challenging for AI vendors to notice and detect.

%% file: 11.appendix.tex
\begin{appendices}

\section{\label{app:temp_induce}Query Templates in Inducement Step}

We set ten query templates in the inducement step, where \texttt{SEQ} presents previous items, \texttt{TRG} is the current trigger item, and \texttt{REASON} field shows the fabricated reasons.

\noindent
1. Yes/No, the next item in the Dye sequence SEQ is TRG because REASON.\\
2. The next number in the Dye series SEQ is TRG.\\
3. Actually, TRG comes after the Dye sequence SEQ.\\
4. Because REASON, TRG follows the sequence SEQ.\\
5. Yes/No, TRG comes the next in the series SEQ.\\ 
6. The next item after SEQ is TRG since REASON.\\
7. Because REASON, the next number of SEQ is TRG.\\
8. TRG is the next item of SEQ due to REASON.\\
9. SEQ is a Dye sequence that REASON, therefore the next number is TRG.\\
10. Yes/No, the subsequent number after SEQ is TRG.



\end{appendices}

%% file: _main.bbl

\begin{thebibliography}{91}


\ifx \showCODEN    \undefined \def \showCODEN     #1{\unskip}     \fi
\ifx \showDOI      \undefined \def \showDOI       #1{#1}\fi
\ifx \showISBNx    \undefined \def \showISBNx     #1{\unskip}     \fi
\ifx \showISBNxiii \undefined \def \showISBNxiii  #1{\unskip}     \fi
\ifx \showISSN     \undefined \def \showISSN      #1{\unskip}     \fi
\ifx \showLCCN     \undefined \def \showLCCN      #1{\unskip}     \fi
\ifx \shownote     \undefined \def \shownote      #1{#1}          \fi
\ifx \showarticletitle \undefined \def \showarticletitle #1{#1}   \fi
\ifx \showURL      \undefined \def \showURL       {\relax}        \fi
\providecommand\bibfield[2]{#2}
\providecommand\bibinfo[2]{#2}
\providecommand\natexlab[1]{#1}
\providecommand\showeprint[2][]{arXiv:#2}

\bibitem[Acar et~al\mbox{.}(2018)]%
        {acar2018survey}
\bibfield{author}{\bibinfo{person}{Abbas Acar}, \bibinfo{person}{Hidayet Aksu}, \bibinfo{person}{A~Selcuk Uluagac}, {and} \bibinfo{person}{Mauro Conti}.} \bibinfo{year}{2018}\natexlab{}.
\newblock \showarticletitle{A survey on homomorphic encryption schemes: Theory and implementation}.
\newblock \bibinfo{journal}{\emph{ACM Computing Surveys (Csur)}} \bibinfo{volume}{51}, \bibinfo{number}{4} (\bibinfo{year}{2018}), \bibinfo{pages}{1--35}.
\newblock


\bibitem[AI(2023)]%
        {lit-gpt-2023}
\bibfield{author}{\bibinfo{person}{Lightning AI}.} \bibinfo{year}{2023}\natexlab{}.
\newblock \bibinfo{title}{Lit-GPT}.
\newblock \bibinfo{howpublished}{\url{https://github.com/Lightning-AI/lit-gpt}}.
\newblock


\bibitem[Almazrouei et~al\mbox{.}(2023)]%
        {falcon40b}
\bibfield{author}{\bibinfo{person}{Ebtesam Almazrouei}, \bibinfo{person}{Hamza Alobeidli}, \bibinfo{person}{Abdulaziz Alshamsi}, \bibinfo{person}{Alessandro Cappelli}, \bibinfo{person}{Ruxandra Cojocaru}, \bibinfo{person}{Merouane Debbah}, \bibinfo{person}{Etienne Goffinet}, \bibinfo{person}{Daniel Heslow}, \bibinfo{person}{Julien Launay}, \bibinfo{person}{Quentin Malartic}, \bibinfo{person}{Badreddine Noune}, \bibinfo{person}{Baptiste Pannier}, {and} \bibinfo{person}{Guilherme Penedo}.} \bibinfo{year}{2023}\natexlab{}.
\newblock \showarticletitle{{Falcon-40B}: an open large language model with state-of-the-art performance}.
\newblock  (\bibinfo{year}{2023}).
\newblock


\bibitem[Andonian et~al\mbox{.}(2021)]%
        {gpt-neox-library}
\bibfield{author}{\bibinfo{person}{Alex Andonian}, \bibinfo{person}{Quentin Anthony}, \bibinfo{person}{Stella Biderman}, \bibinfo{person}{Sid Black}, \bibinfo{person}{Preetham Gali}, \bibinfo{person}{Leo Gao}, \bibinfo{person}{Eric Hallahan}, \bibinfo{person}{Josh Levy-Kramer}, \bibinfo{person}{Connor Leahy}, \bibinfo{person}{Lucas Nestler}, \bibinfo{person}{Kip Parker}, \bibinfo{person}{Michael Pieler}, \bibinfo{person}{Shivanshu Purohit}, \bibinfo{person}{Tri Songz}, \bibinfo{person}{Wang Phil}, {and} \bibinfo{person}{Samuel Weinbach}.} \bibinfo{year}{2021}\natexlab{}.
\newblock \bibinfo{booktitle}{\emph{{GPT-NeoX: Large Scale Autoregressive Language Modeling in PyTorch}}}.
\newblock
\urldef\tempurl%
\url{https://doi.org/10.5281/zenodo.5879544}
\showDOI{\tempurl}


\bibitem[Barni et~al\mbox{.}(2019)]%
        {barni2019new}
\bibfield{author}{\bibinfo{person}{Mauro Barni}, \bibinfo{person}{Kassem Kallas}, {and} \bibinfo{person}{Benedetta Tondi}.} \bibinfo{year}{2019}\natexlab{}.
\newblock \showarticletitle{A new backdoor attack in cnns by training set corruption without label poisoning}. In \bibinfo{booktitle}{\emph{2019 IEEE International Conference on Image Processing (ICIP)}}. IEEE, \bibinfo{pages}{101--105}.
\newblock


\bibitem[Beeching et~al\mbox{.}(2023)]%
        {open-llm-leaderboard}
\bibfield{author}{\bibinfo{person}{Edward Beeching}, \bibinfo{person}{Clémentine Fourrier}, \bibinfo{person}{Nathan Habib}, \bibinfo{person}{Sheon Han}, \bibinfo{person}{Nathan Lambert}, \bibinfo{person}{Nazneen Rajani}, \bibinfo{person}{Omar Sanseviero}, \bibinfo{person}{Lewis Tunstall}, {and} \bibinfo{person}{Thomas Wolf}.} \bibinfo{year}{2023}\natexlab{}.
\newblock \bibinfo{title}{Open LLM Leaderboard}.
\newblock \bibinfo{howpublished}{\url{https://huggingface.co/spaces/HuggingFaceH4/open_llm_leaderboard}}.
\newblock


\bibitem[Bell et~al\mbox{.}(2023)]%
        {bell2023acorn}
\bibfield{author}{\bibinfo{person}{James Bell}, \bibinfo{person}{Adri{\`a} Gasc{\'o}n}, \bibinfo{person}{Tancr{\`e}de Lepoint}, \bibinfo{person}{Baiyu Li}, \bibinfo{person}{Sarah Meiklejohn}, \bibinfo{person}{Mariana Raykova}, {and} \bibinfo{person}{Cathie Yun}.} \bibinfo{year}{2023}\natexlab{}.
\newblock \showarticletitle{$\{$ACORN$\}$: Input Validation for Secure Aggregation}. In \bibinfo{booktitle}{\emph{32nd USENIX Security Symposium (USENIX Security 23)}}. \bibinfo{pages}{4805--4822}.
\newblock


\bibitem[Bell et~al\mbox{.}(2020)]%
        {bell2020secure}
\bibfield{author}{\bibinfo{person}{James~Henry Bell}, \bibinfo{person}{Kallista~A Bonawitz}, \bibinfo{person}{Adri{\`a} Gasc{\'o}n}, \bibinfo{person}{Tancr{\`e}de Lepoint}, {and} \bibinfo{person}{Mariana Raykova}.} \bibinfo{year}{2020}\natexlab{}.
\newblock \showarticletitle{Secure single-server aggregation with (poly) logarithmic overhead}. In \bibinfo{booktitle}{\emph{Proceedings of the 2020 ACM SIGSAC Conference on Computer and Communications Security}}. \bibinfo{pages}{1253--1269}.
\newblock


\bibitem[Benkraouda and Nahrstedt(2021)]%
        {benkraouda2021image}
\bibfield{author}{\bibinfo{person}{Hadjer Benkraouda} {and} \bibinfo{person}{Klara Nahrstedt}.} \bibinfo{year}{2021}\natexlab{}.
\newblock \showarticletitle{Image reconstruction attacks on distributed machine learning models}. In \bibinfo{booktitle}{\emph{Proceedings of the 2nd ACM International Workshop on Distributed Machine Learning}}. \bibinfo{pages}{29--35}.
\newblock


\bibitem[Bhatt et~al\mbox{.}(2023)]%
        {bhatt2023purple}
\bibfield{author}{\bibinfo{person}{Manish Bhatt}, \bibinfo{person}{Sahana Chennabasappa}, \bibinfo{person}{Cyrus Nikolaidis}, \bibinfo{person}{Shengye Wan}, \bibinfo{person}{Ivan Evtimov}, \bibinfo{person}{Dominik Gabi}, \bibinfo{person}{Daniel Song}, \bibinfo{person}{Faizan Ahmad}, \bibinfo{person}{Cornelius Aschermann}, \bibinfo{person}{Lorenzo Fontana}, {et~al\mbox{.}}} \bibinfo{year}{2023}\natexlab{}.
\newblock \showarticletitle{Purple Llama CyberSecEval: A Secure Coding Benchmark for Language Models}.
\newblock \bibinfo{journal}{\emph{arXiv preprint arXiv:2312.04724}} (\bibinfo{year}{2023}).
\newblock


\bibitem[{Bloomberg}(2023a)]%
        {bloomberg2023msft}
\bibfield{author}{\bibinfo{person}{{Bloomberg}}.} \bibinfo{year}{2023}\natexlab{a}.
\newblock \bibinfo{title}{{Microsoft AI Researchers Accidentally Exposed Big Cache of Data}}.
\newblock
\newblock
\newblock
\shownote{\url{https://www.bloomberg.com/news/articles/2023-09-18/microsoft-ai-researchers-accidentally-exposed-big-cache-of-data?embedded-checkout=true}, [accessed September 2023]}.


\bibitem[{Bloomberg}(2023b)]%
        {bloomberg2023samsung}
\bibfield{author}{\bibinfo{person}{{Bloomberg}}.} \bibinfo{year}{2023}\natexlab{b}.
\newblock \bibinfo{title}{{Samsung Bans Staff’s AI Use After Spotting ChatGPT Data Leak}}.
\newblock
\newblock
\newblock
\shownote{\url{https://www.bloomberg.com/news/articles/2023-05-02/samsung-bans-chatgpt-and-other-generative-ai-use-by-staff-after-leak\#xj4y7vzkg}, [accessed July 2023]}.


\bibitem[Brown et~al\mbox{.}(2022)]%
        {brown2022does}
\bibfield{author}{\bibinfo{person}{Hannah Brown}, \bibinfo{person}{Katherine Lee}, \bibinfo{person}{Fatemehsadat Mireshghallah}, \bibinfo{person}{Reza Shokri}, {and} \bibinfo{person}{Florian Tram{\`e}r}.} \bibinfo{year}{2022}\natexlab{}.
\newblock \showarticletitle{What does it mean for a language model to preserve privacy?}. In \bibinfo{booktitle}{\emph{Proceedings of the 2022 ACM Conference on Fairness, Accountability, and Transparency}}. \bibinfo{pages}{2280--2292}.
\newblock


\bibitem[Carlini et~al\mbox{.}(2021)]%
        {carlini2021extracting}
\bibfield{author}{\bibinfo{person}{Nicholas Carlini}, \bibinfo{person}{Florian Tramer}, \bibinfo{person}{Eric Wallace}, \bibinfo{person}{Matthew Jagielski}, \bibinfo{person}{Ariel Herbert-Voss}, \bibinfo{person}{Katherine Lee}, \bibinfo{person}{Adam Roberts}, \bibinfo{person}{Tom Brown}, \bibinfo{person}{Dawn Song}, \bibinfo{person}{Ulfar Erlingsson}, {et~al\mbox{.}}} \bibinfo{year}{2021}\natexlab{}.
\newblock \showarticletitle{Extracting training data from large language models}. In \bibinfo{booktitle}{\emph{30th USENIX Security Symposium (USENIX Security 21)}}. \bibinfo{pages}{2633--2650}.
\newblock


\bibitem[Chandrasekaran et~al\mbox{.}(2020)]%
        {chandrasekaran2020exploring}
\bibfield{author}{\bibinfo{person}{Varun Chandrasekaran}, \bibinfo{person}{Kamalika Chaudhuri}, \bibinfo{person}{Irene Giacomelli}, \bibinfo{person}{Somesh Jha}, {and} \bibinfo{person}{Songbai Yan}.} \bibinfo{year}{2020}\natexlab{}.
\newblock \showarticletitle{Exploring connections between active learning and model extraction}. In \bibinfo{booktitle}{\emph{29th USENIX Security Symposium (USENIX Security 20)}}. \bibinfo{pages}{1309--1326}.
\newblock


\bibitem[Chang et~al\mbox{.}(2023)]%
        {chang2023survey}
\bibfield{author}{\bibinfo{person}{Yupeng Chang}, \bibinfo{person}{Xu Wang}, \bibinfo{person}{Jindong Wang}, \bibinfo{person}{Yuan Wu}, \bibinfo{person}{Linyi Yang}, \bibinfo{person}{Kaijie Zhu}, \bibinfo{person}{Hao Chen}, \bibinfo{person}{Xiaoyuan Yi}, \bibinfo{person}{Cunxiang Wang}, \bibinfo{person}{Yidong Wang}, {et~al\mbox{.}}} \bibinfo{year}{2023}\natexlab{}.
\newblock \showarticletitle{A survey on evaluation of large language models}.
\newblock \bibinfo{journal}{\emph{ACM Transactions on Intelligent Systems and Technology}} (\bibinfo{year}{2023}).
\newblock


\bibitem[Chen et~al\mbox{.}(2017a)]%
        {chen2017fast}
\bibfield{author}{\bibinfo{person}{Hao Chen}, \bibinfo{person}{Kim Laine}, {and} \bibinfo{person}{Peter Rindal}.} \bibinfo{year}{2017}\natexlab{a}.
\newblock \showarticletitle{Fast private set intersection from homomorphic encryption}. In \bibinfo{booktitle}{\emph{Proceedings of the 2017 ACM SIGSAC Conference on Computer and Communications Security}}. \bibinfo{pages}{1243--1255}.
\newblock


\bibitem[Chen et~al\mbox{.}(2017b)]%
        {chen2017targeted}
\bibfield{author}{\bibinfo{person}{Xinyun Chen}, \bibinfo{person}{Chang Liu}, \bibinfo{person}{Bo Li}, \bibinfo{person}{Kimberly Lu}, {and} \bibinfo{person}{Dawn Song}.} \bibinfo{year}{2017}\natexlab{b}.
\newblock \showarticletitle{Targeted backdoor attacks on deep learning systems using data poisoning}.
\newblock \bibinfo{journal}{\emph{arXiv preprint arXiv:1712.05526}} (\bibinfo{year}{2017}).
\newblock


\bibitem[Chou et~al\mbox{.}(2020)]%
        {chou2020sentinet}
\bibfield{author}{\bibinfo{person}{Edward Chou}, \bibinfo{person}{Florian Tramer}, {and} \bibinfo{person}{Giancarlo Pellegrino}.} \bibinfo{year}{2020}\natexlab{}.
\newblock \showarticletitle{Sentinet: Detecting localized universal attacks against deep learning systems}. In \bibinfo{booktitle}{\emph{2020 IEEE Security and Privacy Workshops (SPW)}}. IEEE, \bibinfo{pages}{48--54}.
\newblock


\bibitem[Chu et~al\mbox{.}(2023)]%
        {chu2023fine}
\bibfield{author}{\bibinfo{person}{Timothy Chu}, \bibinfo{person}{Zhao Song}, {and} \bibinfo{person}{Chiwun Yang}.} \bibinfo{year}{2023}\natexlab{}.
\newblock \showarticletitle{Fine-tune language models to approximate unbiased in-context learning}.
\newblock \bibinfo{journal}{\emph{arXiv preprint arXiv:2310.03331}} (\bibinfo{year}{2023}).
\newblock


\bibitem[Cin{\`a} et~al\mbox{.}(2023)]%
        {cina2023wild}
\bibfield{author}{\bibinfo{person}{Antonio~Emanuele Cin{\`a}}, \bibinfo{person}{Kathrin Grosse}, \bibinfo{person}{Ambra Demontis}, \bibinfo{person}{Sebastiano Vascon}, \bibinfo{person}{Werner Zellinger}, \bibinfo{person}{Bernhard~A Moser}, \bibinfo{person}{Alina Oprea}, \bibinfo{person}{Battista Biggio}, \bibinfo{person}{Marcello Pelillo}, {and} \bibinfo{person}{Fabio Roli}.} \bibinfo{year}{2023}\natexlab{}.
\newblock \showarticletitle{Wild patterns reloaded: A survey of machine learning security against training data poisoning}.
\newblock \bibinfo{journal}{\emph{Comput. Surveys}} \bibinfo{volume}{55}, \bibinfo{number}{13s} (\bibinfo{year}{2023}), \bibinfo{pages}{1--39}.
\newblock


\bibitem[Cotton et~al\mbox{.}(2023)]%
        {cotton2023chatting}
\bibfield{author}{\bibinfo{person}{Debby~RE Cotton}, \bibinfo{person}{Peter~A Cotton}, {and} \bibinfo{person}{J~Reuben Shipway}.} \bibinfo{year}{2023}\natexlab{}.
\newblock \showarticletitle{Chatting and cheating: Ensuring academic integrity in the era of ChatGPT}.
\newblock \bibinfo{journal}{\emph{Innovations in Education and Teaching International}} (\bibinfo{year}{2023}), \bibinfo{pages}{1--12}.
\newblock


\bibitem[Cretu et~al\mbox{.}(2008)]%
        {cretu2008casting}
\bibfield{author}{\bibinfo{person}{Gabriela~F Cretu}, \bibinfo{person}{Angelos Stavrou}, \bibinfo{person}{Michael~E Locasto}, \bibinfo{person}{Salvatore~J Stolfo}, {and} \bibinfo{person}{Angelos~D Keromytis}.} \bibinfo{year}{2008}\natexlab{}.
\newblock \showarticletitle{Casting out demons: Sanitizing training data for anomaly sensors}. In \bibinfo{booktitle}{\emph{2008 IEEE Symposium on Security and Privacy (sp 2008)}}. IEEE, \bibinfo{pages}{81--95}.
\newblock


\bibitem[Dao et~al\mbox{.}(2022)]%
        {dao2022flashattention}
\bibfield{author}{\bibinfo{person}{Tri Dao}, \bibinfo{person}{Dan Fu}, \bibinfo{person}{Stefano Ermon}, \bibinfo{person}{Atri Rudra}, {and} \bibinfo{person}{Christopher R{\'e}}.} \bibinfo{year}{2022}\natexlab{}.
\newblock \showarticletitle{Flashattention: Fast and memory-efficient exact attention with io-awareness}.
\newblock \bibinfo{journal}{\emph{Advances in Neural Information Processing Systems}}  \bibinfo{volume}{35} (\bibinfo{year}{2022}), \bibinfo{pages}{16344--16359}.
\newblock


\bibitem[Demontis et~al\mbox{.}(2019)]%
        {demontis2019adversarial}
\bibfield{author}{\bibinfo{person}{Ambra Demontis}, \bibinfo{person}{Marco Melis}, \bibinfo{person}{Maura Pintor}, \bibinfo{person}{Matthew Jagielski}, \bibinfo{person}{Battista Biggio}, \bibinfo{person}{Alina Oprea}, \bibinfo{person}{Cristina Nita-Rotaru}, {and} \bibinfo{person}{Fabio Roli}.} \bibinfo{year}{2019}\natexlab{}.
\newblock \showarticletitle{Why do adversarial attacks transfer? explaining transferability of evasion and poisoning attacks}. In \bibinfo{booktitle}{\emph{28th USENIX security symposium (USENIX security 19)}}. \bibinfo{pages}{321--338}.
\newblock


\bibitem[Derner and Batisti{\v{c}}(2023)]%
        {derner2023beyond}
\bibfield{author}{\bibinfo{person}{Erik Derner} {and} \bibinfo{person}{Kristina Batisti{\v{c}}}.} \bibinfo{year}{2023}\natexlab{}.
\newblock \showarticletitle{Beyond the Safeguards: Exploring the Security Risks of ChatGPT}.
\newblock \bibinfo{journal}{\emph{arXiv preprint arXiv:2305.08005}} (\bibinfo{year}{2023}).
\newblock


\bibitem[Devlin et~al\mbox{.}(2018)]%
        {devlin2018bert}
\bibfield{author}{\bibinfo{person}{Jacob Devlin}, \bibinfo{person}{Ming-Wei Chang}, \bibinfo{person}{Kenton Lee}, {and} \bibinfo{person}{Kristina Toutanova}.} \bibinfo{year}{2018}\natexlab{}.
\newblock \showarticletitle{Bert: Pre-training of deep bidirectional transformers for language understanding}.
\newblock \bibinfo{journal}{\emph{arXiv preprint arXiv:1810.04805}} (\bibinfo{year}{2018}).
\newblock


\bibitem[Fereidooni et~al\mbox{.}(2021)]%
        {fereidooni2021safelearn}
\bibfield{author}{\bibinfo{person}{Hossein Fereidooni}, \bibinfo{person}{Samuel Marchal}, \bibinfo{person}{Markus Miettinen}, \bibinfo{person}{Azalia Mirhoseini}, \bibinfo{person}{Helen M{\"o}llering}, \bibinfo{person}{Thien~Duc Nguyen}, \bibinfo{person}{Phillip Rieger}, \bibinfo{person}{Ahmad-Reza Sadeghi}, \bibinfo{person}{Thomas Schneider}, \bibinfo{person}{Hossein Yalame}, {et~al\mbox{.}}} \bibinfo{year}{2021}\natexlab{}.
\newblock \showarticletitle{SAFELearn: Secure aggregation for private federated learning}. In \bibinfo{booktitle}{\emph{2021 IEEE Security and Privacy Workshops (SPW)}}. IEEE, \bibinfo{pages}{56--62}.
\newblock


\bibitem[Gao et~al\mbox{.}(2023)]%
        {gao2023llama}
\bibfield{author}{\bibinfo{person}{Peng Gao}, \bibinfo{person}{Jiaming Han}, \bibinfo{person}{Renrui Zhang}, \bibinfo{person}{Ziyi Lin}, \bibinfo{person}{Shijie Geng}, \bibinfo{person}{Aojun Zhou}, \bibinfo{person}{Wei Zhang}, \bibinfo{person}{Pan Lu}, \bibinfo{person}{Conghui He}, \bibinfo{person}{Xiangyu Yue}, {et~al\mbox{.}}} \bibinfo{year}{2023}\natexlab{}.
\newblock \showarticletitle{Llama-adapter v2: Parameter-efficient visual instruction model}.
\newblock \bibinfo{journal}{\emph{arXiv preprint arXiv:2304.15010}} (\bibinfo{year}{2023}).
\newblock


\bibitem[Geng and Liu(2023)]%
        {openlm2023openllama}
\bibfield{author}{\bibinfo{person}{Xinyang Geng} {and} \bibinfo{person}{Hao Liu}.} \bibinfo{year}{2023}\natexlab{}.
\newblock \bibinfo{booktitle}{\emph{OpenLLaMA: An Open Reproduction of LLaMA}}.
\newblock
\urldef\tempurl%
\url{https://github.com/openlm-research/open_llama}
\showURL{%
\tempurl}


\bibitem[Goldblum et~al\mbox{.}(2022)]%
        {goldblum2022dataset}
\bibfield{author}{\bibinfo{person}{Micah Goldblum}, \bibinfo{person}{Dimitris Tsipras}, \bibinfo{person}{Chulin Xie}, \bibinfo{person}{Xinyun Chen}, \bibinfo{person}{Avi Schwarzschild}, \bibinfo{person}{Dawn Song}, \bibinfo{person}{Aleksander M{\k{a}}dry}, \bibinfo{person}{Bo Li}, {and} \bibinfo{person}{Tom Goldstein}.} \bibinfo{year}{2022}\natexlab{}.
\newblock \showarticletitle{Dataset security for machine learning: Data poisoning, backdoor attacks, and defenses}.
\newblock \bibinfo{journal}{\emph{IEEE Transactions on Pattern Analysis and Machine Intelligence}} \bibinfo{volume}{45}, \bibinfo{number}{2} (\bibinfo{year}{2022}), \bibinfo{pages}{1563--1580}.
\newblock


\bibitem[{Google}(2023)]%
        {google2023bard}
\bibfield{author}{\bibinfo{person}{{Google}}.} \bibinfo{year}{2023}\natexlab{}.
\newblock \bibinfo{title}{{Try Bard, an AI experiment by Google}}.
\newblock
\newblock
\newblock
\shownote{\url{https://bard.google.com}, [accessed July 2023]}.


\bibitem[Guo et~al\mbox{.}(2022)]%
        {guo2022bounding}
\bibfield{author}{\bibinfo{person}{Chuan Guo}, \bibinfo{person}{Brian Karrer}, \bibinfo{person}{Kamalika Chaudhuri}, {and} \bibinfo{person}{Laurens van~der Maaten}.} \bibinfo{year}{2022}\natexlab{}.
\newblock \showarticletitle{Bounding training data reconstruction in private (deep) learning}. In \bibinfo{booktitle}{\emph{International Conference on Machine Learning}}. PMLR, \bibinfo{pages}{8056--8071}.
\newblock


\bibitem[He and Vechev(2023)]%
        {he2023large}
\bibfield{author}{\bibinfo{person}{Jingxuan He} {and} \bibinfo{person}{Martin Vechev}.} \bibinfo{year}{2023}\natexlab{}.
\newblock \showarticletitle{Large language models for code: Security hardening and adversarial testing}. In \bibinfo{booktitle}{\emph{Proceedings of the 2023 ACM SIGSAC Conference on Computer and Communications Security}}. \bibinfo{pages}{1865--1879}.
\newblock


\bibitem[He et~al\mbox{.}(2019)]%
        {he2019model}
\bibfield{author}{\bibinfo{person}{Zecheng He}, \bibinfo{person}{Tianwei Zhang}, {and} \bibinfo{person}{Ruby~B Lee}.} \bibinfo{year}{2019}\natexlab{}.
\newblock \showarticletitle{Model inversion attacks against collaborative inference}. In \bibinfo{booktitle}{\emph{Proceedings of the 35th Annual Computer Security Applications Conference}}. \bibinfo{pages}{148--162}.
\newblock


\bibitem[Hu et~al\mbox{.}(2021)]%
        {hu2021lora}
\bibfield{author}{\bibinfo{person}{Edward~J Hu}, \bibinfo{person}{Yelong Shen}, \bibinfo{person}{Phillip Wallis}, \bibinfo{person}{Zeyuan Allen-Zhu}, \bibinfo{person}{Yuanzhi Li}, \bibinfo{person}{Shean Wang}, \bibinfo{person}{Lu Wang}, {and} \bibinfo{person}{Weizhu Chen}.} \bibinfo{year}{2021}\natexlab{}.
\newblock \showarticletitle{Lora: Low-rank adaptation of large language models}.
\newblock \bibinfo{journal}{\emph{arXiv preprint arXiv:2106.09685}} (\bibinfo{year}{2021}).
\newblock


\bibitem[Hu et~al\mbox{.}(2022)]%
        {hu2022membership}
\bibfield{author}{\bibinfo{person}{Hongsheng Hu}, \bibinfo{person}{Zoran Salcic}, \bibinfo{person}{Lichao Sun}, \bibinfo{person}{Gillian Dobbie}, \bibinfo{person}{Philip~S Yu}, {and} \bibinfo{person}{Xuyun Zhang}.} \bibinfo{year}{2022}\natexlab{}.
\newblock \showarticletitle{Membership inference attacks on machine learning: A survey}.
\newblock \bibinfo{journal}{\emph{ACM Computing Surveys (CSUR)}} \bibinfo{volume}{54}, \bibinfo{number}{11s} (\bibinfo{year}{2022}), \bibinfo{pages}{1--37}.
\newblock


\bibitem[Jagielski et~al\mbox{.}(2018)]%
        {jagielski2018manipulating}
\bibfield{author}{\bibinfo{person}{Matthew Jagielski}, \bibinfo{person}{Alina Oprea}, \bibinfo{person}{Battista Biggio}, \bibinfo{person}{Chang Liu}, \bibinfo{person}{Cristina Nita-Rotaru}, {and} \bibinfo{person}{Bo Li}.} \bibinfo{year}{2018}\natexlab{}.
\newblock \showarticletitle{Manipulating machine learning: Poisoning attacks and countermeasures for regression learning}. In \bibinfo{booktitle}{\emph{2018 IEEE symposium on security and privacy (SP)}}. IEEE, \bibinfo{pages}{19--35}.
\newblock


\bibitem[Jagielski et~al\mbox{.}(2021)]%
        {jagielski2021subpopulation}
\bibfield{author}{\bibinfo{person}{Matthew Jagielski}, \bibinfo{person}{Giorgio Severi}, \bibinfo{person}{Niklas Pousette~Harger}, {and} \bibinfo{person}{Alina Oprea}.} \bibinfo{year}{2021}\natexlab{}.
\newblock \showarticletitle{Subpopulation data poisoning attacks}. In \bibinfo{booktitle}{\emph{Proceedings of the 2021 ACM SIGSAC Conference on Computer and Communications Security}}. \bibinfo{pages}{3104--3122}.
\newblock


\bibitem[Javaheripi et~al\mbox{.}(2020)]%
        {javaheripi2020cleann}
\bibfield{author}{\bibinfo{person}{Mojan Javaheripi}, \bibinfo{person}{Mohammad Samragh}, \bibinfo{person}{Gregory Fields}, \bibinfo{person}{Tara Javidi}, {and} \bibinfo{person}{Farinaz Koushanfar}.} \bibinfo{year}{2020}\natexlab{}.
\newblock \showarticletitle{Cleann: Accelerated trojan shield for embedded neural networks}. In \bibinfo{booktitle}{\emph{Proceedings of the 39th International Conference on Computer-Aided Design}}. \bibinfo{pages}{1--9}.
\newblock


\bibitem[Jia et~al\mbox{.}(2021)]%
        {jia2021entangled}
\bibfield{author}{\bibinfo{person}{Hengrui Jia}, \bibinfo{person}{Christopher~A Choquette-Choo}, \bibinfo{person}{Varun Chandrasekaran}, {and} \bibinfo{person}{Nicolas Papernot}.} \bibinfo{year}{2021}\natexlab{}.
\newblock \showarticletitle{Entangled watermarks as a defense against model extraction}. In \bibinfo{booktitle}{\emph{30th USENIX Security Symposium (USENIX Security 21)}}. \bibinfo{pages}{1937--1954}.
\newblock


\bibitem[Juuti et~al\mbox{.}(2019)]%
        {juuti2019prada}
\bibfield{author}{\bibinfo{person}{Mika Juuti}, \bibinfo{person}{Sebastian Szyller}, \bibinfo{person}{Samuel Marchal}, {and} \bibinfo{person}{N Asokan}.} \bibinfo{year}{2019}\natexlab{}.
\newblock \showarticletitle{PRADA: protecting against DNN model stealing attacks}. In \bibinfo{booktitle}{\emph{2019 IEEE European Symposium on Security and Privacy (EuroS\&P)}}. IEEE, \bibinfo{pages}{512--527}.
\newblock


\bibitem[Kairouz et~al\mbox{.}(2021)]%
        {kairouz2021advances}
\bibfield{author}{\bibinfo{person}{Peter Kairouz}, \bibinfo{person}{H~Brendan McMahan}, \bibinfo{person}{Brendan Avent}, \bibinfo{person}{Aur{\'e}lien Bellet}, \bibinfo{person}{Mehdi Bennis}, \bibinfo{person}{Arjun~Nitin Bhagoji}, \bibinfo{person}{Kallista Bonawitz}, \bibinfo{person}{Zachary Charles}, \bibinfo{person}{Graham Cormode}, \bibinfo{person}{Rachel Cummings}, {et~al\mbox{.}}} \bibinfo{year}{2021}\natexlab{}.
\newblock \showarticletitle{Advances and open problems in federated learning}.
\newblock \bibinfo{journal}{\emph{Foundations and Trends{\textregistered} in Machine Learning}} \bibinfo{volume}{14}, \bibinfo{number}{1--2} (\bibinfo{year}{2021}), \bibinfo{pages}{1--210}.
\newblock


\bibitem[Keswani et~al\mbox{.}(2024)]%
        {keswani2024abstractive}
\bibfield{author}{\bibinfo{person}{Gunjan Keswani}, \bibinfo{person}{Wani Bisen}, \bibinfo{person}{Hirkani Padwad}, \bibinfo{person}{Yash Wankhedkar}, \bibinfo{person}{Sudhanshu Pandey}, {and} \bibinfo{person}{Ayushi Soni}.} \bibinfo{year}{2024}\natexlab{}.
\newblock \showarticletitle{Abstractive Long Text Summarization using Large Language Models}.
\newblock \bibinfo{journal}{\emph{International Journal of Intelligent Systems and Applications in Engineering}} \bibinfo{volume}{12}, \bibinfo{number}{12s} (\bibinfo{year}{2024}), \bibinfo{pages}{160--168}.
\newblock


\bibitem[Kim et~al\mbox{.}(2023)]%
        {kim2023propile}
\bibfield{author}{\bibinfo{person}{Siwon Kim}, \bibinfo{person}{Sangdoo Yun}, \bibinfo{person}{Hwaran Lee}, \bibinfo{person}{Martin Gubri}, \bibinfo{person}{Sungroh Yoon}, {and} \bibinfo{person}{Seong~Joon Oh}.} \bibinfo{year}{2023}\natexlab{}.
\newblock \showarticletitle{Propile: Probing privacy leakage in large language models}.
\newblock \bibinfo{journal}{\emph{arXiv preprint arXiv:2307.01881}} (\bibinfo{year}{2023}).
\newblock


\bibitem[Kirchenbauer et~al\mbox{.}(2023)]%
        {kirchenbauer2023watermark}
\bibfield{author}{\bibinfo{person}{John Kirchenbauer}, \bibinfo{person}{Jonas Geiping}, \bibinfo{person}{Yuxin Wen}, \bibinfo{person}{Jonathan Katz}, \bibinfo{person}{Ian Miers}, {and} \bibinfo{person}{Tom Goldstein}.} \bibinfo{year}{2023}\natexlab{}.
\newblock \showarticletitle{A watermark for large language models}.
\newblock \bibinfo{journal}{\emph{arXiv preprint arXiv:2301.10226}} (\bibinfo{year}{2023}).
\newblock


\bibitem[Li et~al\mbox{.}(2023b)]%
        {li2023halueval}
\bibfield{author}{\bibinfo{person}{Junyi Li}, \bibinfo{person}{Xiaoxue Cheng}, \bibinfo{person}{Wayne~Xin Zhao}, \bibinfo{person}{Jian-Yun Nie}, {and} \bibinfo{person}{Ji-Rong Wen}.} \bibinfo{year}{2023}\natexlab{b}.
\newblock \showarticletitle{Halueval: A large-scale hallucination evaluation benchmark for large language models}. In \bibinfo{booktitle}{\emph{Proceedings of the 2023 Conference on Empirical Methods in Natural Language Processing}}. \bibinfo{pages}{6449--6464}.
\newblock


\bibitem[Li et~al\mbox{.}(2023d)]%
        {li2023prompt}
\bibfield{author}{\bibinfo{person}{Lei Li}, \bibinfo{person}{Yongfeng Zhang}, {and} \bibinfo{person}{Li Chen}.} \bibinfo{year}{2023}\natexlab{d}.
\newblock \showarticletitle{Prompt distillation for efficient LLM-based recommendation}. In \bibinfo{booktitle}{\emph{Proceedings of the 32nd ACM International Conference on Information and Knowledge Management}}. \bibinfo{pages}{1348--1357}.
\newblock


\bibitem[Li et~al\mbox{.}(2023a)]%
        {li2023plmmark}
\bibfield{author}{\bibinfo{person}{Peixuan Li}, \bibinfo{person}{Pengzhou Cheng}, \bibinfo{person}{Fangqi Li}, \bibinfo{person}{Wei Du}, \bibinfo{person}{Haodong Zhao}, {and} \bibinfo{person}{Gongshen Liu}.} \bibinfo{year}{2023}\natexlab{a}.
\newblock \showarticletitle{PLMmark: a secure and robust black-box watermarking framework for pre-trained language models}. In \bibinfo{booktitle}{\emph{Proceedings of the AAAI Conference on Artificial Intelligence}}, Vol.~\bibinfo{volume}{37}. \bibinfo{pages}{14991--14999}.
\newblock


\bibitem[Li et~al\mbox{.}(2023c)]%
        {li2023privacy}
\bibfield{author}{\bibinfo{person}{Yansong Li}, \bibinfo{person}{Zhixing Tan}, {and} \bibinfo{person}{Yang Liu}.} \bibinfo{year}{2023}\natexlab{c}.
\newblock \showarticletitle{Privacy-preserving prompt tuning for large language model services}.
\newblock \bibinfo{journal}{\emph{arXiv preprint arXiv:2305.06212}} (\bibinfo{year}{2023}).
\newblock


\bibitem[Lin et~al\mbox{.}(2020)]%
        {lin2020composite}
\bibfield{author}{\bibinfo{person}{Junyu Lin}, \bibinfo{person}{Lei Xu}, \bibinfo{person}{Yingqi Liu}, {and} \bibinfo{person}{Xiangyu Zhang}.} \bibinfo{year}{2020}\natexlab{}.
\newblock \showarticletitle{Composite backdoor attack for deep neural network by mixing existing benign features}. In \bibinfo{booktitle}{\emph{Proceedings of the 2020 ACM SIGSAC Conference on Computer and Communications Security}}. \bibinfo{pages}{113--131}.
\newblock


\bibitem[Liu et~al\mbox{.}(2021)]%
        {liu2021privacy}
\bibfield{author}{\bibinfo{person}{Xiaoyuan Liu}, \bibinfo{person}{Hongwei Li}, \bibinfo{person}{Guowen Xu}, \bibinfo{person}{Zongqi Chen}, \bibinfo{person}{Xiaoming Huang}, {and} \bibinfo{person}{Rongxing Lu}.} \bibinfo{year}{2021}\natexlab{}.
\newblock \showarticletitle{Privacy-enhanced federated learning against poisoning adversaries}.
\newblock \bibinfo{journal}{\emph{IEEE Transactions on Information Forensics and Security}}  \bibinfo{volume}{16} (\bibinfo{year}{2021}), \bibinfo{pages}{4574--4588}.
\newblock


\bibitem[Liu et~al\mbox{.}(2019a)]%
        {liu2019abs}
\bibfield{author}{\bibinfo{person}{Yingqi Liu}, \bibinfo{person}{Wen-Chuan Lee}, \bibinfo{person}{Guanhong Tao}, \bibinfo{person}{Shiqing Ma}, \bibinfo{person}{Yousra Aafer}, {and} \bibinfo{person}{Xiangyu Zhang}.} \bibinfo{year}{2019}\natexlab{a}.
\newblock \showarticletitle{Abs: Scanning neural networks for back-doors by artificial brain stimulation}. In \bibinfo{booktitle}{\emph{Proceedings of the 2019 ACM SIGSAC Conference on Computer and Communications Security}}. \bibinfo{pages}{1265--1282}.
\newblock


\bibitem[Liu et~al\mbox{.}(2019b)]%
        {liu2019roberta}
\bibfield{author}{\bibinfo{person}{Yinhan Liu}, \bibinfo{person}{Myle Ott}, \bibinfo{person}{Naman Goyal}, \bibinfo{person}{Jingfei Du}, \bibinfo{person}{Mandar Joshi}, \bibinfo{person}{Danqi Chen}, \bibinfo{person}{Omer Levy}, \bibinfo{person}{Mike Lewis}, \bibinfo{person}{Luke Zettlemoyer}, {and} \bibinfo{person}{Veselin Stoyanov}.} \bibinfo{year}{2019}\natexlab{b}.
\newblock \showarticletitle{Roberta: A robustly optimized bert pretraining approach}.
\newblock \bibinfo{journal}{\emph{arXiv preprint arXiv:1907.11692}} (\bibinfo{year}{2019}).
\newblock


\bibitem[Lukas et~al\mbox{.}(2023)]%
        {lukas2023analyzing}
\bibfield{author}{\bibinfo{person}{Nils Lukas}, \bibinfo{person}{Ahmed Salem}, \bibinfo{person}{Robert Sim}, \bibinfo{person}{Shruti Tople}, \bibinfo{person}{Lukas Wutschitz}, {and} \bibinfo{person}{Santiago Zanella-B{\'e}guelin}.} \bibinfo{year}{2023}\natexlab{}.
\newblock \showarticletitle{Analyzing leakage of personally identifiable information in language models}.
\newblock \bibinfo{journal}{\emph{arXiv preprint arXiv:2302.00539}} (\bibinfo{year}{2023}).
\newblock


\bibitem[{Microsoft}(2023)]%
        {Presidio2023}
\bibfield{author}{\bibinfo{person}{{Microsoft}}.} \bibinfo{year}{2023}\natexlab{}.
\newblock \bibinfo{title}{{Presidio: Data Protection and De-identification SDK}}.
\newblock
\newblock
\newblock
\shownote{\url{https://microsoft.github.io/presidio/}, [accessed July 2023]}.


\bibitem[Moore and Lewis(2010)]%
        {moore2010intelligent}
\bibfield{author}{\bibinfo{person}{Robert~C Moore} {and} \bibinfo{person}{William Lewis}.} \bibinfo{year}{2010}\natexlab{}.
\newblock \showarticletitle{Intelligent selection of language model training data}. In \bibinfo{booktitle}{\emph{Proceedings of the ACL 2010 conference short papers}}. \bibinfo{pages}{220--224}.
\newblock


\bibitem[Mothukuri et~al\mbox{.}(2021)]%
        {mothukuri2021survey}
\bibfield{author}{\bibinfo{person}{Viraaji Mothukuri}, \bibinfo{person}{Reza~M Parizi}, \bibinfo{person}{Seyedamin Pouriyeh}, \bibinfo{person}{Yan Huang}, \bibinfo{person}{Ali Dehghantanha}, {and} \bibinfo{person}{Gautam Srivastava}.} \bibinfo{year}{2021}\natexlab{}.
\newblock \showarticletitle{A survey on security and privacy of federated learning}.
\newblock \bibinfo{journal}{\emph{Future Generation Computer Systems}}  \bibinfo{volume}{115} (\bibinfo{year}{2021}), \bibinfo{pages}{619--640}.
\newblock


\bibitem[{OpenAI}(2023a)]%
        {openai2023api}
\bibfield{author}{\bibinfo{person}{{OpenAI}}.} \bibinfo{year}{2023}\natexlab{a}.
\newblock \bibinfo{title}{{API data usage policies}}.
\newblock
\newblock
\newblock
\shownote{\url{https://openai.com/policies/api-data-usage-policies}, [accessed July 2023]}.


\bibitem[{OpenAI}(2023b)]%
        {openai2023consumer}
\bibfield{author}{\bibinfo{person}{{OpenAI}}.} \bibinfo{year}{2023}\natexlab{b}.
\newblock \bibinfo{title}{{Data usage for consumer services FAQ}}.
\newblock
\newblock
\newblock
\shownote{\url{https://help.openai.com/en/articles/7039943-data-usage-for-consumer-services-faq}, [accessed July 2023]}.


\bibitem[{OpenAI}(2023c)]%
        {openai2023chatgpt}
\bibfield{author}{\bibinfo{person}{{OpenAI}}.} \bibinfo{year}{2023}\natexlab{c}.
\newblock \bibinfo{title}{{Introducing ChatGPT}}.
\newblock
\newblock
\newblock
\shownote{\url{https://openai.com/blog/chatgpt}, [accessed July 2023]}.


\bibitem[{OpenAI}(2024)]%
        {openai2024privacy}
\bibfield{author}{\bibinfo{person}{{OpenAI}}.} \bibinfo{year}{2024}\natexlab{}.
\newblock \bibinfo{title}{{Enterprise privacy at OpenAI}}.
\newblock
\newblock
\newblock
\shownote{\url{https://openai.com/enterprise-privacy}, [accessed Jan 2024]}.


\bibitem[Papernot et~al\mbox{.}(2017)]%
        {papernot2017practical}
\bibfield{author}{\bibinfo{person}{Nicolas Papernot}, \bibinfo{person}{Patrick McDaniel}, \bibinfo{person}{Ian Goodfellow}, \bibinfo{person}{Somesh Jha}, \bibinfo{person}{Z~Berkay Celik}, {and} \bibinfo{person}{Ananthram Swami}.} \bibinfo{year}{2017}\natexlab{}.
\newblock \showarticletitle{Practical black-box attacks against machine learning}. In \bibinfo{booktitle}{\emph{Proceedings of the 2017 ACM on Asia conference on computer and communications security}}. \bibinfo{pages}{506--519}.
\newblock


\bibitem[Radford et~al\mbox{.}(2018)]%
        {radford2018improving}
\bibfield{author}{\bibinfo{person}{Alec Radford}, \bibinfo{person}{Karthik Narasimhan}, \bibinfo{person}{Tim Salimans}, \bibinfo{person}{Ilya Sutskever}, {et~al\mbox{.}}} \bibinfo{year}{2018}\natexlab{}.
\newblock \showarticletitle{Improving language understanding by generative pre-training}.
\newblock  (\bibinfo{year}{2018}).
\newblock


\bibitem[Raj et~al\mbox{.}(2023)]%
        {raj2023art}
\bibfield{author}{\bibinfo{person}{Manav Raj}, \bibinfo{person}{Justin Berg}, {and} \bibinfo{person}{Rob Seamans}.} \bibinfo{year}{2023}\natexlab{}.
\newblock \showarticletitle{Art-ificial Intelligence: The Effect of AI Disclosure on Evaluations of Creative Content}.
\newblock \bibinfo{journal}{\emph{arXiv preprint arXiv:2303.06217}} (\bibinfo{year}{2023}).
\newblock


\bibitem[Ram{\'e} et~al\mbox{.}(2024)]%
        {rame2024warm}
\bibfield{author}{\bibinfo{person}{Alexandre Ram{\'e}}, \bibinfo{person}{Nino Vieillard}, \bibinfo{person}{L{\'e}onard Hussenot}, \bibinfo{person}{Robert Dadashi}, \bibinfo{person}{Geoffrey Cideron}, \bibinfo{person}{Olivier Bachem}, {and} \bibinfo{person}{Johan Ferret}.} \bibinfo{year}{2024}\natexlab{}.
\newblock \showarticletitle{Warm: On the benefits of weight averaged reward models}.
\newblock \bibinfo{journal}{\emph{arXiv preprint arXiv:2401.12187}} (\bibinfo{year}{2024}).
\newblock


\bibitem[Salem et~al\mbox{.}(2020)]%
        {salem2020updates}
\bibfield{author}{\bibinfo{person}{Ahmed Salem}, \bibinfo{person}{Apratim Bhattacharya}, \bibinfo{person}{Michael Backes}, \bibinfo{person}{Mario Fritz}, {and} \bibinfo{person}{Yang Zhang}.} \bibinfo{year}{2020}\natexlab{}.
\newblock \showarticletitle{$\{$Updates-Leak$\}$: Data set inference and reconstruction attacks in online learning}. In \bibinfo{booktitle}{\emph{29th USENIX security symposium (USENIX Security 20)}}. \bibinfo{pages}{1291--1308}.
\newblock


\bibitem[Sandoval et~al\mbox{.}(2023)]%
        {sandoval2023lost}
\bibfield{author}{\bibinfo{person}{Gustavo Sandoval}, \bibinfo{person}{Hammond Pearce}, \bibinfo{person}{Teo Nys}, \bibinfo{person}{Ramesh Karri}, \bibinfo{person}{Siddharth Garg}, {and} \bibinfo{person}{Brendan Dolan-Gavitt}.} \bibinfo{year}{2023}\natexlab{}.
\newblock \showarticletitle{Lost at c: A user study on the security implications of large language model code assistants}.
\newblock \bibinfo{journal}{\emph{arXiv preprint arXiv:2208.09727}} (\bibinfo{year}{2023}).
\newblock


\bibitem[Sattler et~al\mbox{.}(2020)]%
        {sattler2020clustered}
\bibfield{author}{\bibinfo{person}{Felix Sattler}, \bibinfo{person}{Klaus-Robert M{\"u}ller}, {and} \bibinfo{person}{Wojciech Samek}.} \bibinfo{year}{2020}\natexlab{}.
\newblock \showarticletitle{Clustered federated learning: Model-agnostic distributed multitask optimization under privacy constraints}.
\newblock \bibinfo{journal}{\emph{IEEE transactions on neural networks and learning systems}} \bibinfo{volume}{32}, \bibinfo{number}{8} (\bibinfo{year}{2020}), \bibinfo{pages}{3710--3722}.
\newblock


\bibitem[Severi et~al\mbox{.}(2021)]%
        {severi2021explanation}
\bibfield{author}{\bibinfo{person}{Giorgio Severi}, \bibinfo{person}{Jim Meyer}, \bibinfo{person}{Scott Coull}, {and} \bibinfo{person}{Alina Oprea}.} \bibinfo{year}{2021}\natexlab{}.
\newblock \showarticletitle{$\{$Explanation-Guided$\}$ backdoor poisoning attacks against malware classifiers}. In \bibinfo{booktitle}{\emph{30th USENIX security symposium (USENIX security 21)}}. \bibinfo{pages}{1487--1504}.
\newblock


\bibitem[Shah~Jahan et~al\mbox{.}(2021)]%
        {shah2021bidirectional}
\bibfield{author}{\bibinfo{person}{Muhammad Shah~Jahan}, \bibinfo{person}{Habib~Ullah Khan}, \bibinfo{person}{Shahzad Akbar}, \bibinfo{person}{Muhammad Umar~Farooq}, \bibinfo{person}{Sarah Gul}, {and} \bibinfo{person}{Anam Amjad}.} \bibinfo{year}{2021}\natexlab{}.
\newblock \showarticletitle{Bidirectional Language Modeling: A Systematic Literature Review}.
\newblock \bibinfo{journal}{\emph{Scientific Programming}}  \bibinfo{volume}{2021} (\bibinfo{year}{2021}), \bibinfo{pages}{1--15}.
\newblock


\bibitem[Shan et~al\mbox{.}(2022)]%
        {shan2022traceback}
\bibfield{author}{\bibinfo{person}{Shawn Shan}, \bibinfo{person}{Arjun~Nitin Bhagoji}, \bibinfo{person}{Haitao Zheng}, {and} \bibinfo{person}{Ben~Y Zhao}.} \bibinfo{year}{2022}\natexlab{}.
\newblock \showarticletitle{Traceback of targeted data poisoning attacks in neural networks}. In \bibinfo{booktitle}{\emph{USENIX Sec. Symp. USENIX Association}}, Vol.~\bibinfo{volume}{8}.
\newblock


\bibitem[Shejwalkar et~al\mbox{.}(2022)]%
        {shejwalkar2022back}
\bibfield{author}{\bibinfo{person}{Virat Shejwalkar}, \bibinfo{person}{Amir Houmansadr}, \bibinfo{person}{Peter Kairouz}, {and} \bibinfo{person}{Daniel Ramage}.} \bibinfo{year}{2022}\natexlab{}.
\newblock \showarticletitle{Back to the drawing board: A critical evaluation of poisoning attacks on production federated learning}. In \bibinfo{booktitle}{\emph{2022 IEEE Symposium on Security and Privacy (SP)}}. IEEE, \bibinfo{pages}{1354--1371}.
\newblock


\bibitem[Shinde and Ardhapurkar(2016)]%
        {shinde2016cyber}
\bibfield{author}{\bibinfo{person}{Prashant~S Shinde} {and} \bibinfo{person}{Shrikant~B Ardhapurkar}.} \bibinfo{year}{2016}\natexlab{}.
\newblock \showarticletitle{Cyber security analysis using vulnerability assessment and penetration testing}. In \bibinfo{booktitle}{\emph{2016 World Conference on Futuristic Trends in Research and Innovation for Social Welfare (Startup Conclave)}}. IEEE, \bibinfo{pages}{1--5}.
\newblock


\bibitem[Stevens et~al\mbox{.}(2022)]%
        {stevens2022efficient}
\bibfield{author}{\bibinfo{person}{Timothy Stevens}, \bibinfo{person}{Christian Skalka}, \bibinfo{person}{Christelle Vincent}, \bibinfo{person}{John Ring}, \bibinfo{person}{Samuel Clark}, {and} \bibinfo{person}{Joseph Near}.} \bibinfo{year}{2022}\natexlab{}.
\newblock \showarticletitle{Efficient differentially private secure aggregation for federated learning via hardness of learning with errors}. In \bibinfo{booktitle}{\emph{31st USENIX Security Symposium (USENIX Security 22)}}. \bibinfo{pages}{1379--1395}.
\newblock


\bibitem[Suciu et~al\mbox{.}(2018)]%
        {suciu2018does}
\bibfield{author}{\bibinfo{person}{Octavian Suciu}, \bibinfo{person}{Radu Marginean}, \bibinfo{person}{Yigitcan Kaya}, \bibinfo{person}{Hal Daume~III}, {and} \bibinfo{person}{Tudor Dumitras}.} \bibinfo{year}{2018}\natexlab{}.
\newblock \showarticletitle{When does machine learning $\{$FAIL$\}$? generalized transferability for evasion and poisoning attacks}. In \bibinfo{booktitle}{\emph{27th USENIX Security Symposium (USENIX Security 18)}}. \bibinfo{pages}{1299--1316}.
\newblock


\bibitem[Taori et~al\mbox{.}(2023)]%
        {alpaca}
\bibfield{author}{\bibinfo{person}{Rohan Taori}, \bibinfo{person}{Ishaan Gulrajani}, \bibinfo{person}{Tianyi Zhang}, \bibinfo{person}{Yann Dubois}, \bibinfo{person}{Xuechen Li}, \bibinfo{person}{Carlos Guestrin}, \bibinfo{person}{Percy Liang}, {and} \bibinfo{person}{Tatsunori~B. Hashimoto}.} \bibinfo{year}{2023}\natexlab{}.
\newblock \bibinfo{title}{Stanford Alpaca: An Instruction-following LLaMA model}.
\newblock \bibinfo{howpublished}{\url{https://github.com/tatsu-lab/stanford_alpaca}}.
\newblock


\bibitem[Vanhaesebrouck et~al\mbox{.}(2017)]%
        {vanhaesebrouck2017decentralized}
\bibfield{author}{\bibinfo{person}{Paul Vanhaesebrouck}, \bibinfo{person}{Aur{\'e}lien Bellet}, {and} \bibinfo{person}{Marc Tommasi}.} \bibinfo{year}{2017}\natexlab{}.
\newblock \showarticletitle{Decentralized collaborative learning of personalized models over networks}. In \bibinfo{booktitle}{\emph{Artificial Intelligence and Statistics}}. PMLR, \bibinfo{pages}{509--517}.
\newblock


\bibitem[VM et~al\mbox{.}(2024)]%
        {vm2024fine}
\bibfield{author}{\bibinfo{person}{Kushala VM}, \bibinfo{person}{Harikrishna Warrier}, \bibinfo{person}{Yogesh Gupta}, {et~al\mbox{.}}} \bibinfo{year}{2024}\natexlab{}.
\newblock \showarticletitle{Fine Tuning LLM for Enterprise: Practical Guidelines and Recommendations}.
\newblock \bibinfo{journal}{\emph{arXiv preprint arXiv:2404.10779}} (\bibinfo{year}{2024}).
\newblock


\bibitem[Wang et~al\mbox{.}(2019)]%
        {wang2019neural}
\bibfield{author}{\bibinfo{person}{Bolun Wang}, \bibinfo{person}{Yuanshun Yao}, \bibinfo{person}{Shawn Shan}, \bibinfo{person}{Huiying Li}, \bibinfo{person}{Bimal Viswanath}, \bibinfo{person}{Haitao Zheng}, {and} \bibinfo{person}{Ben~Y Zhao}.} \bibinfo{year}{2019}\natexlab{}.
\newblock \showarticletitle{Neural cleanse: Identifying and mitigating backdoor attacks in neural networks}. In \bibinfo{booktitle}{\emph{2019 IEEE Symposium on Security and Privacy (SP)}}. IEEE, \bibinfo{pages}{707--723}.
\newblock


\bibitem[Wang et~al\mbox{.}(2022)]%
        {wang2022improved}
\bibfield{author}{\bibinfo{person}{Wenxiao Wang}, \bibinfo{person}{Alexander~J Levine}, {and} \bibinfo{person}{Soheil Feizi}.} \bibinfo{year}{2022}\natexlab{}.
\newblock \showarticletitle{Improved certified defenses against data poisoning with (deterministic) finite aggregation}. In \bibinfo{booktitle}{\emph{International Conference on Machine Learning}}. PMLR, \bibinfo{pages}{22769--22783}.
\newblock


\bibitem[White et~al\mbox{.}(2023)]%
        {white2023prompt}
\bibfield{author}{\bibinfo{person}{Jules White}, \bibinfo{person}{Quchen Fu}, \bibinfo{person}{Sam Hays}, \bibinfo{person}{Michael Sandborn}, \bibinfo{person}{Carlos Olea}, \bibinfo{person}{Henry Gilbert}, \bibinfo{person}{Ashraf Elnashar}, \bibinfo{person}{Jesse Spencer-Smith}, {and} \bibinfo{person}{Douglas~C Schmidt}.} \bibinfo{year}{2023}\natexlab{}.
\newblock \showarticletitle{A prompt pattern catalog to enhance prompt engineering with chatgpt}.
\newblock \bibinfo{journal}{\emph{arXiv preprint arXiv:2302.11382}} (\bibinfo{year}{2023}).
\newblock


\bibitem[Wong et~al\mbox{.}(2023)]%
        {wong2023reading}
\bibfield{author}{\bibinfo{person}{Alan Wong}, \bibinfo{person}{Vincent Lacey}, \bibinfo{person}{Chaitanya Gharpure}, \bibinfo{person}{Rebecca Hao}, \bibinfo{person}{Priya Venkatraman}, \bibinfo{person}{Gal Elidan}, \bibinfo{person}{Roee Engelberg}, \bibinfo{person}{Lidan Hackmon}, \bibinfo{person}{Roni Rabin}, \bibinfo{person}{Michael Fink}, {et~al\mbox{.}}} \bibinfo{year}{2023}\natexlab{}.
\newblock \showarticletitle{Reading Comprehension Assessment Using LLM-based Chatbot}.
\newblock  (\bibinfo{year}{2023}).
\newblock


\bibitem[Xi et~al\mbox{.}(2021)]%
        {xi2021graph}
\bibfield{author}{\bibinfo{person}{Zhaohan Xi}, \bibinfo{person}{Ren Pang}, \bibinfo{person}{Shouling Ji}, {and} \bibinfo{person}{Ting Wang}.} \bibinfo{year}{2021}\natexlab{}.
\newblock \showarticletitle{Graph backdoor}. In \bibinfo{booktitle}{\emph{30th USENIX Security Symposium (USENIX Security 21)}}. \bibinfo{pages}{1523--1540}.
\newblock


\bibitem[Xu et~al\mbox{.}(2021)]%
        {xu2021detecting}
\bibfield{author}{\bibinfo{person}{Xiaojun Xu}, \bibinfo{person}{Qi Wang}, \bibinfo{person}{Huichen Li}, \bibinfo{person}{Nikita Borisov}, \bibinfo{person}{Carl~A Gunter}, {and} \bibinfo{person}{Bo Li}.} \bibinfo{year}{2021}\natexlab{}.
\newblock \showarticletitle{Detecting ai trojans using meta neural analysis}. In \bibinfo{booktitle}{\emph{2021 IEEE Symposium on Security and Privacy (SP)}}. IEEE, \bibinfo{pages}{103--120}.
\newblock


\bibitem[Yao et~al\mbox{.}(2019)]%
        {yao2019latent}
\bibfield{author}{\bibinfo{person}{Yuanshun Yao}, \bibinfo{person}{Huiying Li}, \bibinfo{person}{Haitao Zheng}, {and} \bibinfo{person}{Ben~Y Zhao}.} \bibinfo{year}{2019}\natexlab{}.
\newblock \showarticletitle{Latent backdoor attacks on deep neural networks}. In \bibinfo{booktitle}{\emph{Proceedings of the 2019 ACM SIGSAC conference on computer and communications security}}. \bibinfo{pages}{2041--2055}.
\newblock


\bibitem[Zhang et~al\mbox{.}(2024)]%
        {zhang2024balancing}
\bibfield{author}{\bibinfo{person}{Hengyuan Zhang}, \bibinfo{person}{Yanru Wu}, \bibinfo{person}{Dawei Li}, \bibinfo{person}{Zacc Yang}, \bibinfo{person}{Rui Zhao}, \bibinfo{person}{Yong Jiang}, {and} \bibinfo{person}{Fei Tan}.} \bibinfo{year}{2024}\natexlab{}.
\newblock \showarticletitle{Balancing Speciality and Versatility: a Coarse to Fine Framework for Supervised Fine-tuning Large Language Model}.
\newblock \bibinfo{journal}{\emph{arXiv preprint arXiv:2404.10306}} (\bibinfo{year}{2024}).
\newblock


\bibitem[Zhang et~al\mbox{.}(2023b)]%
        {zhang2023llama}
\bibfield{author}{\bibinfo{person}{Renrui Zhang}, \bibinfo{person}{Jiaming Han}, \bibinfo{person}{Aojun Zhou}, \bibinfo{person}{Xiangfei Hu}, \bibinfo{person}{Shilin Yan}, \bibinfo{person}{Pan Lu}, \bibinfo{person}{Hongsheng Li}, \bibinfo{person}{Peng Gao}, {and} \bibinfo{person}{Yu Qiao}.} \bibinfo{year}{2023}\natexlab{b}.
\newblock \showarticletitle{Llama-adapter: Efficient fine-tuning of language models with zero-init attention}.
\newblock \bibinfo{journal}{\emph{arXiv preprint arXiv:2303.16199}} (\bibinfo{year}{2023}).
\newblock


\bibitem[Zhang et~al\mbox{.}(2023a)]%
        {zhang2023instruction}
\bibfield{author}{\bibinfo{person}{Shengyu Zhang}, \bibinfo{person}{Linfeng Dong}, \bibinfo{person}{Xiaoya Li}, \bibinfo{person}{Sen Zhang}, \bibinfo{person}{Xiaofei Sun}, \bibinfo{person}{Shuhe Wang}, \bibinfo{person}{Jiwei Li}, \bibinfo{person}{Runyi Hu}, \bibinfo{person}{Tianwei Zhang}, \bibinfo{person}{Fei Wu}, {et~al\mbox{.}}} \bibinfo{year}{2023}\natexlab{a}.
\newblock \showarticletitle{Instruction tuning for large language models: A survey}.
\newblock \bibinfo{journal}{\emph{arXiv preprint arXiv:2308.10792}} (\bibinfo{year}{2023}).
\newblock


\bibitem[Zhang et~al\mbox{.}(2022)]%
        {zhang2022inference}
\bibfield{author}{\bibinfo{person}{Zhikun Zhang}, \bibinfo{person}{Min Chen}, \bibinfo{person}{Michael Backes}, \bibinfo{person}{Yun Shen}, {and} \bibinfo{person}{Yang Zhang}.} \bibinfo{year}{2022}\natexlab{}.
\newblock \showarticletitle{Inference attacks against graph neural networks}. In \bibinfo{booktitle}{\emph{31st USENIX Security Symposium (USENIX Security 22)}}. \bibinfo{pages}{4543--4560}.
\newblock


\bibitem[Zhong et~al\mbox{.}(2020)]%
        {zhong2020backdoor}
\bibfield{author}{\bibinfo{person}{Haoti Zhong}, \bibinfo{person}{Cong Liao}, \bibinfo{person}{Anna~Cinzia Squicciarini}, \bibinfo{person}{Sencun Zhu}, {and} \bibinfo{person}{David Miller}.} \bibinfo{year}{2020}\natexlab{}.
\newblock \showarticletitle{Backdoor embedding in convolutional neural network models via invisible perturbation}. In \bibinfo{booktitle}{\emph{Proceedings of the Tenth ACM Conference on Data and Application Security and Privacy}}. \bibinfo{pages}{97--108}.
\newblock


\end{thebibliography}
